\documentclass[aps,prb,twocolumn,superscriptaddress,groupedaddress,showpacs]{revtex4-1}  

\usepackage{graphicx}  
\usepackage{dcolumn}   
\usepackage{bm}        
\usepackage{amssymb}   
\usepackage{amsmath}
\usepackage{natbib}
\usepackage{float}
\usepackage{xcolor}
\usepackage{soul}

\usepackage{comment}

\begin{document}

\title{Ising-nematic order in the bilinear-biquadratic model for the iron pnictides}
\author{Patricia Bilbao Ergueta}
\author{Andriy H. Nevidomskyy}
\affiliation{Department of Physics and Astronomy, Rice University, Houston,
TX 77005}

\let\oldhat\hat
\renewcommand{\vec}[1]{\mathbf{#1}}
\renewcommand{\hat}[1]{\oldhat{\mathbf{#1}}}

\newcommand{\beq}{\begin{equation}}
\newcommand{\eeq}{\end{equation}}
\newcommand{\vS}{\mathbf{S}}
\newcommand{\vQ}{\mathbf{Q}}
\newcommand{\vq}{\mathbf{q}}
\newcommand{\vk}{\mathbf{k}}
\newcommand{\hx}{\hat{x}}
\newcommand{\hy}{\hat{y}}
\newcommand{\hz}{\hat{z}}
\newcommand{\mean}[1]{\langle#1\rangle}
\newcommand{\ud}{\mathrm{d}}
\newcommand{\an}[1]{\textcolor{blue}{#1}}
\newcommand{\ahn}[1]{\textcolor{red}{#1}}

\begin{abstract}
Motivated by the recent inelastic neutron scattering (INS) measurements in the iron pnictides which show a strong anisotropy of spin excitations in directions perpendicular and parallel to the ordering wave-vector even above the magnetic transition temperature $T_N$, we study the frustrated Heisenberg model with a biquadratic spin-spin exchange interaction. Using the Dyson-Maleev (DM) representation, which proves appropriate for all temperature regimes, we find that the spin-spin dynamical structure factors are in excellent agreement with experiment, exhibiting breaking of the $C_4$ symmetry even into the paramagnetic region $T_N<T<T_{\sigma}$, which we refer to as the Ising-nematic phase. In addition to the Heisenberg spin interaction, we include the biquadratic coupling $-K (\vS_i\cdot \vS_j)^2$ and study its effect on the dynamical temperature range $T_{\sigma}-T_N$ of the Ising-nematic phase. We find that this range reduces dramatically when even small values of the interlayer exchange $J_c$ and biquadratic coupling $K$ are included. 
To supplement our analysis, we benchmark the results obtained using full decoupling in the DM method against those from different non-linear spin-wave theories, including the recently developed generalized spin-wave theory (GSWT), and find good qualitative agreement among the different theoretical approaches as well as experiment for both the spin-wave dispersions and the dynamical structure factors.
\end{abstract}

\maketitle

\section{\label{sec:intro}Introduction}

The recent discovery of iron-based high-$T_c$ superconductors \cite{Kamihara2008,Ren2008} and the antiferromagnetically ordered nature of their parent compounds \cite{DelaCruz2008} sheds more light on the importance of understanding the electronic correlations and the magnetic excitations in these materials, especially due to the similarities between their phase diagram and that of the cuprates. Like the cuprates, the parent compounds of the iron pnictides exhibit an antiferromagnetic ground state below the N\'eel ordering temperature $T_N$. However instead of the regular N\'eel phase, pnictides order into a columnar antiferromagnet (CAF) with the ordering wave-vector $\vQ=\left(\pi,0\right)$ or $\left(0,\pi\right)$~\cite{DelaCruz2008}. 
The transition to the long-range magnetic order in the pnictides takes place in a very close proximity to a structural transition from a tetragonal to an orthorhombic phase below temperature $T_s\geq T_N$. Therefore, it is not \emph{a priori} clear whether the electronic degrees of freedom or the purely elastic lattice response play the primary role.
Recent resistivity $\rho$ measurements under fixed strain $\delta=(a-b)/(a+b)$ ($a$ and $b$ are the in-plane lattice constants) have detected divergent nematic susceptibility $\ud \rho/\ud\delta$, proving that the nematicity is of electronic origin rather than due to an elastic lattice instability\cite{Chu2012}.

Two different mechanisms have been proposed for the electronic nematic ordering: the spin-nematic scenario~\cite{Fang2008, Xu2008, Dai2009, Fernandes2011, Fernandes2012} and the orbital ordering with unequal population of iron $d_{xz}$ and $d_{yz}$ orbitals~\cite{Kruger2009, Lv2009, Lee2009, Chen2009, Lv2010}.
Because the corresponding order parameters are coupled linearly in the Landau free energy, the appearance of one will cause a non-zero expectation value of the other, and disentangling which is the cause and which is the consequence is very difficult~\cite{Fernandes2014}. In this work, we will not attempt to answer this question. Rather, we are interested in the physical signatures of electronic nematic order in the electron spin response.
In particular, we are motivated by the recent inelastic neutron scattering (INS) data on detwinned BaFe$_2$As$_2$ that exhibit a two-fold anisotropy in the spin excitations even above the structural transition temperature $T_s$ in the nominally tetragonal phase~\cite{Lu2014}. 

To address this problem theoretically, we chose an approach in terms of quasi-local moments on Fe sites, following earlier works by other authors~\cite{Si2008,Yildirim2008,Ma2008,Fang2008,Xu2008,Si2009,Dai2009,Uhrig2009}. This strong coupling perspective is motivated by the ``bad metal'' nature of the parent compounds and the proposed proximity to the Mott localization transition~\cite{Si2008,Si2009,Haule2008,Kutepov2010}. Indeed, superconductivity was found to border a Mott insulating phase in alkaline iron selenides $A_{1-x}$Fe$_{2-y}$Se$_2$ (the ``245" family, with $A=$ K, Rb, Cs, or Tl)~\cite{Fang2011,Wang2011,Bao2011,Wang2011a}. 
The Mott insulating ground state has also been identified in the iron oxychalcogenides La$_2$O$_3$Fe$_2$(Se,S)$_2$ [\onlinecite{Zhu2010},\onlinecite{McCabe2014}], $R_2$O$_3$Fe$_2$Se$_2$ (here $R$ = Ce,Pr,Nd, or Sm)~\cite{Ni2011}, and Sr$_2$F$_2$Fe$_2$OS$_2$~[\onlinecite{Zhao2013}]. Further evidence in favor of proximity to the incipient Mott phase comes from the suppression of the Drude peak in optical conductivity measurements on iron pnictides~\cite{Qazilbash2009,Hu2008} and the spectral weight transfer induced by temperature \cite{Yang2009,Boris2009}. Theoretically, it was proposed that Hund's coupling plays a crucial role in enhancing strong electron correlations in the iron pnictides~\cite{Haule2009}, due to the fact that it decouples charge fluctuations in individual orbitals, leading to the orbital-dependent mass renormalization~\cite{Yin2011,Craco2011,Physique2014}. These predictions have been confirmed experimentally e.g. in KFe$_2$As$_2$~[\onlinecite{Yoshida2014}]. As the Coulomb repulsion strength grows, this has been suggested to eventually result in an orbital-selective Mott transition~\cite{Yu2013}, observed in alkaline iron selenides~\cite{Yi2013, Wang2014}.

While some aspects of magnetism can be understood from a weak coupling approach of itinerant electrons with the Fermi surface nesting~\cite{Graser2009,Ran2009,Knolle2010}, the aforementioned studies justify the use of an effective model of localized spins $\vec{S}_i$ on iron sites with nearest ($<\!\!i,j\!\!>$) and next-nearest neighbor ($\ll i,k \gg$) interactions:
\beq
\begin{aligned}
\mathcal{H}_{3D}&=J_1\sum_{<i,j>}\vec{S}_{i}\cdot\vec{S}_{j}- K\sum_{<i,j>}\left(\vec{S}_{i}\cdot\vec{S}_{j}\right)^2 \\
&+J_2\sum_{\ll i,k\gg}\vec{S}_{i}\cdot\vec{S}_{k} + J_c \sum_i \vS_i \cdot \vS_{i + \hz}
\end{aligned}
\label{eq:J1-J2-K}
\eeq
The size of the effective spin $S$ is dictated by the strength of the Hund's coupling relative to the Fe $d$-electron bandwidth and crystal-field splittings. In the parent compounds of iron pnictides, $S=1$ agrees with the integrated spin spectral weight of $\sim3 \mu_B$ per Fe from INS measurements~\cite{Liu2012}.
In the iron chalcogenides, the electron bandwidth is narrower, so that even though the absolute value of the Hund's coupling is similar to that in the iron pnictides ($J_H\sim 0.7$~eV),  its role is more pronounced, resulting in a larger spin $S=2$.

In addition to the Heisenberg spin interaction, the effective Hamiltonian (\ref{eq:J1-J2-K}) also contains the biquadratic spin-spin interactions. The latter are important to correctly capture the dispersion of the spin excitations near the Brillouin zone boundary~\cite{Wysocki2011,Stanek2011,Yu2012} observed via the INS~\cite{Zhao2009, Harriger2011}.
Indeed, without the biquadratic $K$-term, one is forced to adopt an anisotropic nearest neighbor (NN) coupling constant \cite{Zhao2008, Zhao2009,Han2009} $J_{1x}\neq J_{1y}$ even when modeling the neutron spectra above the structural transition~\cite{Harriger2011}, which is unphysical. Moreover, such analysis would predict $J_1$ to be wildly different in two crystallographic directions, with antiferromagnetic $J_{1x}$ and ferromagnetic $J_{1y}$ which are impossible to reconcile even when the small ($\delta \lesssim 1\%$) orthorhombic lattice distortion is taken into account. 
Instead, the inclusion of a biquadratic term in Eq.~(\ref{eq:J1-J2-K}) dynamically generates the anisotropy in the effective NN Heisenberg couplings, in agreement with the experimental results~\cite{Wysocki2011,Stanek2011,Yu2012}. 

In this work, we show that the spin wave dispersion and dynamical spin structure factor from INS measurements~\cite{Harriger2011,Lu2014} can be modeled semi-quantitatively using the effective spin model~(\ref{eq:J1-J2-K}). Moreover, we find that there is a temperature range $T_N < T <T_\sigma$ in the paramagnetic phase with nematic anisotropy of the spin excitations, similar to the recent INS data on BaFe$_{2-x}$Ni$_x$As$_2$ [\onlinecite{Lu2014}]. On a technical level, this work improves significantly upon the earlier work by one of the co-authors~\cite{Yu2012}, which used a simpler decoupling of the biquadratic spin-spin interaction. By contrast, here we employ several more exacting methods to treat the model Eq.~(\ref{eq:J1-J2-K}), including a non-linear spin-wave theory, the Dyson--Maleev spin representation~\cite{Dyson1956,Dyson1956a,Maleev1958}, and the recently developed so-called generalized spin-wave theory (GSWT)~\cite{Muniz2013}.
Last but not the least, we demonstrate that the origin of the biquadratic spin interaction $-K(\vS_i\cdot \vS_j)^2$ can be understood in terms of the multi-orbital nature of the iron pnictides by deriving, at the mean-field level, the connection between the model Eq.~(\ref{eq:J1-J2-K}) and the Kugel--Khomskii Hamiltonian Eq.~(\ref{eq:KK}), which describes the coupled spin and orbital degrees of freedom. Despite the mean-field nature of our derivation (see Section~\ref{sec:model} below), it helps establish an important connection between the orbital mechanism of nematicity and the spin response captured by the $J_1-J_2-K$ model. It also allows one to understand why, in the presence of orbital ordering, the value of the biquadratic  coupling constant $K$ can be relatively large, comparable to the Heisenberg interactions $J_{1,2}$, which is difficult to justify otherwise.

The paper is organized as follows. First, we provide the aforementioned derivation of the effective spin Hamiltonian in Sec.~\ref{sec:model}.
Then, in Sec.~\ref{sec:results}, we summarize our main results and compare them to experiment. We then go on to show the spin-wave dispersions obtained with the different methods in Sec.~\ref{sec:disp}. In Sec.~\ref{sec:struc} we include plots for the low temperature dynamical structure factors and comment on their agreement with experimental results. We analyze the evolution of both the staggered magnetization and the nematic order parameter with temperature in detail in Sec.~\ref{sec:nem} and finally summarize our conclusions in Sec.~\ref{sec:concl}. The details of the different methods used to calculate spin-waves are given in the Appendices for convenience.

\section{\label{sec:model} Effective Spin Model}

In the original Refs.~\onlinecite{Wysocki2011,Stanek2011,Yu2012}, the biquadratic spin-spin interaction in Eq.~(\ref{eq:J1-J2-K}) was introduced heuristically as a higher order spin exchange process, derived for instance as a fourth order perturbation in the Schrieffer--Wolff projection of the Hubbard model~\cite{Mila2000}. 
In this section we show that alternatively, the microscopic origin of the $K$-term can be traced to orbital ordering within the theory of coupled spin and orbital degrees of freedom. The advantage of such an interpretation is that it allows to incorporate the orbital physics into the spin response, since ultimately both orbital and spin degrees of freedom are involved in the electronic nematic phase~\cite{Fernandes2014,Wang2014a}.

To see this explicitly, let us consider as a starting point the  Kugel--Khomskii model~\cite{Kugel1982} formulated for the $d_{xz}$ and $d_{yz}$ orbitals of the iron pnictides following Refs.~\onlinecite{Kruger2009,Lv2009,Chen2009,Singh2009}:
\beq
H\!\! = \!\!J\sum_i\left[(\vS_i\cdot \vS_{i+\hx}+1)\tau_i^a \tau_{i+\hx}^a + (\vS_i\cdot \vS_{i+\hy}+1)\tau_i^b \tau_{i+\hy}^b \right] 
\label{eq:KK}
\eeq
where $\vS_i=1$ is the magnetic moment originating from the Hund's coupled electron spins in Fe $d_{xz}$ and $d_{yz}$ orbitals, and $\{\tau^a,\tau^b\}$ are the pseudospin operators that act in the orbital subspace of $|xz\rangle$ and $|yz\rangle$ states and depend on the directionality of the Fe-Fe bond. This model is complicated to deal with, but for our purposes, a simple mean-field decoupling will suffice, with the resulting ground state energy
\beq
E_{MF}\! = J\left[ (\Gamma_x+1) \langle \tau_i^a \tau_{i+\hx}^a \rangle + (\Gamma_y + 1) \langle \tau_i^b \tau_{i+\hy}^b \rangle \right],
\label{eq:MF}
\eeq
where we have introduced the following notation for the bond expectation values:
\beq
\Gamma_x \equiv \langle \vS_i\cdot\vS_{i+\hx} \rangle,\qquad \Gamma_y \equiv \langle \vS_i\cdot\vS_{i+\hy} \rangle.
\label{eq:Gamma}
\eeq

To proceed, we relate the average of the orbital pseudospin operators to the occupation number of the corresponding orbital: $\mean{\tau^a} = n_{xz}$ and $\mean{\tau^b} = n_{yz}$. This is similar to earlier studies in Refs.~\onlinecite{Lv2009,Chen2009,Singh2009}, who treated $\tau^\alpha$ as Ising degrees of freedom. We point out that $n_{xz}$ and $n_{yz}$ orbitals are occupied by one electron each in the tetragonal phase of the parent compound, so that the total occupancy $n = n_{xz} + n_{yz} = 2$. Then, the correlators $\mean{\tau_i \tau_{i+\gamma}}$ can be written as follows:
\begin{eqnarray}
\left\langle \tau^a_i \tau^a_{i+\hx}\right\rangle &=& \left(n_{xz} + \frac{\delta n}{2}\right)^2 = \frac{n+P+\delta n}{2} \nonumber \\
\left\langle \tau^b_i \tau^b_{i+\hy}\right\rangle &=& \left(n_{yz} + \frac{\delta n}{2}\right)^2 = \frac{n-P+\delta n}{2}
\label{eq:decouple}
\end{eqnarray}
where we have introduced the orbital polarization $P=n_{xz} - n_{yz}$ and $\delta n$ is a phenomenological parameter that accounts for orbital fluctuations. Indeed, $\delta n$ is a measure of the variance in the orbital occupation number $\mean{\tau^a_i \tau^a_{i+\hx}} - \mean{\tau^a_i}^2 = n_{xz} \delta n + \mathcal{O}(\delta n^2) $ and similarly for  the $yz$-orbital.

Minimizing the mean-field energy Eq.~(\ref{eq:MF}) with respect to $P$, we find the expectation value of the orbital polarization
\beq
P = \frac{n+\delta n}{2+\Gamma_x + \Gamma_y} (\Gamma_y - \Gamma_x) \approx (\Gamma_y - \Gamma_x)
\label{eq:P}
\eeq
where the last equality is obtained by noting that \mbox{$\Gamma_x = -\Gamma_y$} in the columnar antiferromagnetic phase and $n+\delta n \approx 2$. We see that the orbital polarization is linearly proportional to the Ising-nematic order parameter ($\Gamma_y - \Gamma_x$), as it should be based on the Landau theory of coupled order parameters, since the bilinear combination $P\cdot(\Gamma_y-\Gamma_x)$ is allowed by symmetry.
Using Eqs. (\ref{eq:decouple}) and (\ref{eq:P}), we can now decouple the orbital degrees of freedom in the Kugel--Khomskii model (\ref{eq:KK}) at the mean-field level, resulting in an effective spin Hamiltonian to linear order in $P$:
\begin{eqnarray}
\label{eq:H_MF}
H_\text{MF}\! &\sim& \!\left(\frac{n+\delta n}{2}\right)^2  J \sum_{\langle i,j  \rangle} \vS_i\cdot \vS_j  \\ 
&-& \!\left(\frac{n+\delta n}{2}\right)^2 J \sum_i (\Gamma_y - \Gamma_x) (\vS_i\cdot\vS_{i+\hy} - \vS_i\cdot\vS_{i+\hx} ) \nonumber
\end{eqnarray}
If we now forget that this mean-field Hamiltonian originated from orbital physics, it is tempting to interpret it as a mean-field approximation to the following spin Hamiltonian:
\begin{eqnarray}
H_\text{spin} &=& J_1 \sum_{\langle i,j  \rangle} \vS_i\cdot \vS_j - \frac{J_1}{2} \sum_{\langle i,j  \rangle} (\vS_i\cdot \vS_j)^2 \nonumber \\
&+& J_1 \sum_i (\vS_i\cdot \vS_{i+\hx})(\vS_i\cdot \vS_{i+\hy})
\label{eq:Hspin}
\end{eqnarray}
with $J_1 = J (n+\delta n)^2/4$.
The first two terms of this effective Hamiltonian are the same as in Eq.~(\ref{eq:J1-J2-K}), with the biquadratic term appearing naturally as a result of integrating out the orbital degrees of freedom. The last term in Eq.~(\ref{eq:Hspin}) involves three spin interactions and similar to the $J_2$ Heisenberg term, favors columnar antiferromagnetic order with $\mean{S_i\cdot S_{i+\hx}} = -\mean{S_i\cdot S_{i+\hy}}$. 
Of course, it is understood that the above argument is not a truly microscopic derivation of the $J_1-J_2-K$ model Eq.~(\ref{eq:J1-J2-K}). Rather, it proves that both Eq.~(\ref{eq:J1-J2-K}) and the Kugel--Khomskii Hamiltonian Eq.~(\ref{eq:KK}) share roughly the same mean-field Hamiltonian described by Eq.~(\ref{eq:Hspin}). Nevertheless, it allows to construct an important link between the orbital nematic order studied by many authors~\cite{Kruger2009, Lv2009, Lee2009, Chen2009, Lv2010, Chen2010, Kontani2011} and the effective spin response of the iron pnictides. It is also important to point out that, within this derivation, one obtains a value of $K=J_1/2$, explaining the relatively large value of $K\approx 0.6 J_1$ necessary to fit the spin wave dispersion from inelastic neutron scattering~\cite{Wysocki2011,Yu2012}.
For completeness, we mention that the possibility of generating a significantly large coupling $|K|$ provided the system has quasi-degenerate orbitals has also been pointed out by Mila and Zhang~\cite{Mila2000} who used higher order perturbation theory to derive the biquadratic spin exchange from the Hubbard model. 
Also, the biquadratic spin-spin interaction can be obtained as a result of the magnetoelastic coupling, provided the lattice has suitable phonon modes, as was proposed for Fe$_{1+y}$Te in Ref.~\onlinecite{Cano2010}.

\section{\label{sec:results} Main results}

At low enough temperatures ($T<T_N$) the iron pnictides exhibit a columnar antiferromagnetic spin-stripe ground state with two energetically degenerate wave-vectors $\vec{Q}=(\pi,0)$ or $(0,\pi)$. Within this region, the staggered magnetization $m_s$ has a finite value. The magnetic transition temperature is always equal or lower than the structural transition temperature $T_s$,
and the INS experiments~\cite{Harriger2011} have found an anisotropy of the spin-wave dispersions above $T_N$. 
Therefore, there is a finite range of temperatures $T_N<T<T_s$ with a nematic anisotropy in the spin response, prompting the researchers to call this a spin Ising-nematic phase~\cite{Fang2008,Xu2008}, following early ideas of spontaneous $Z_2$ symmetry breaking in the frustrated $J_1 - J_2$ model on a square lattice~\cite{Chandra1990}.

\begin{figure}[!t]
\centering
\includegraphics[width=0.4\textwidth]{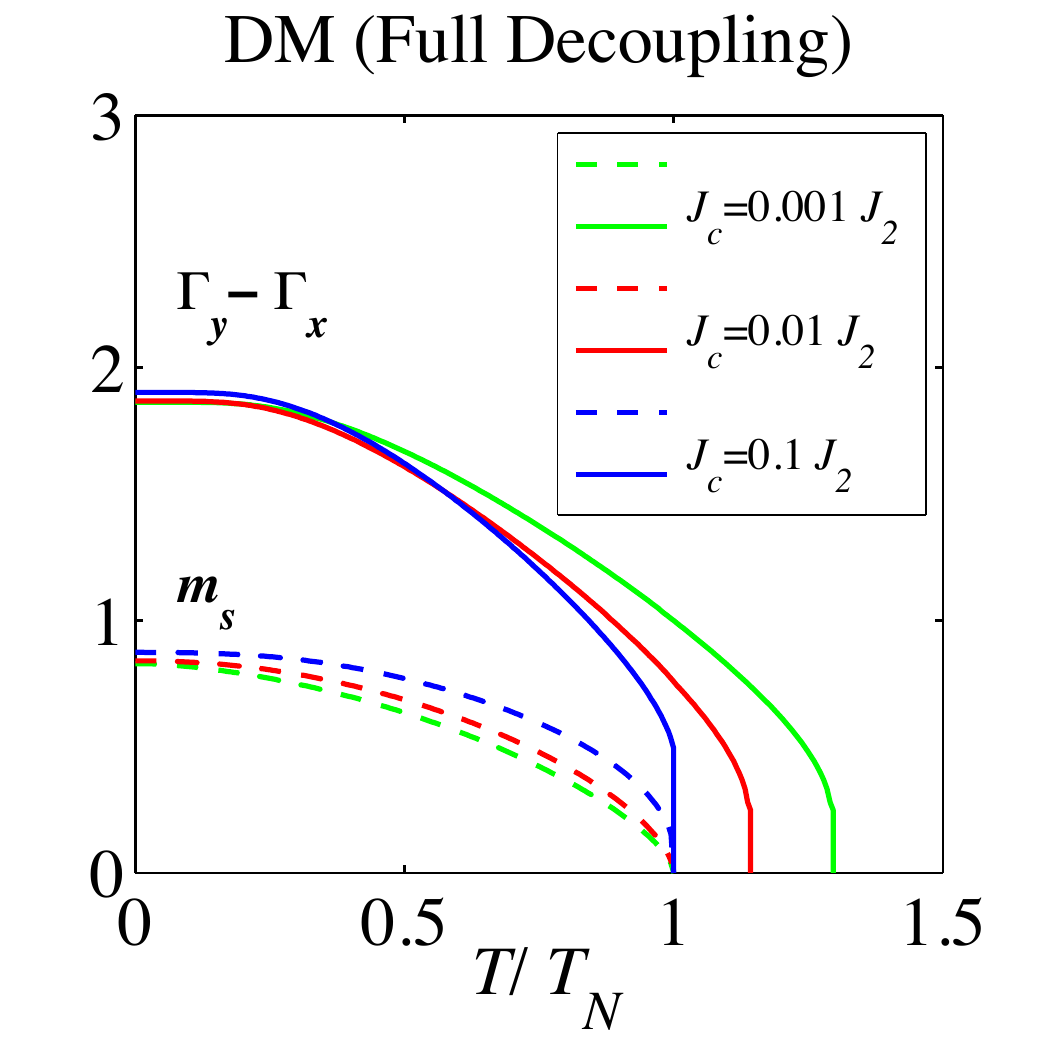}
\caption{(Color online) Evolution of the sublattice magnetization $m_s$ and the nematic order parameter $\left(\Gamma_y-\Gamma_x\right)$ for the case of $S=1$ and $K=0.01J_2$, for different values of the interlayer coupling using the full decoupling of the DM bosons. As expected, $T_N\leq T_{\sigma}$ for all cases although the nematic range of temperatures is negligible except for artificially small values of $J_c$ (corresponding effectively to a quasi two-dimensional model). Note that the discontinuous first order transition observed for the nematicity is just an artifice of the mean-field technique.}
\label{fig:dm_params}
\end{figure}

\begin{figure}[!th]
\centering
\includegraphics[width=0.4\textwidth]{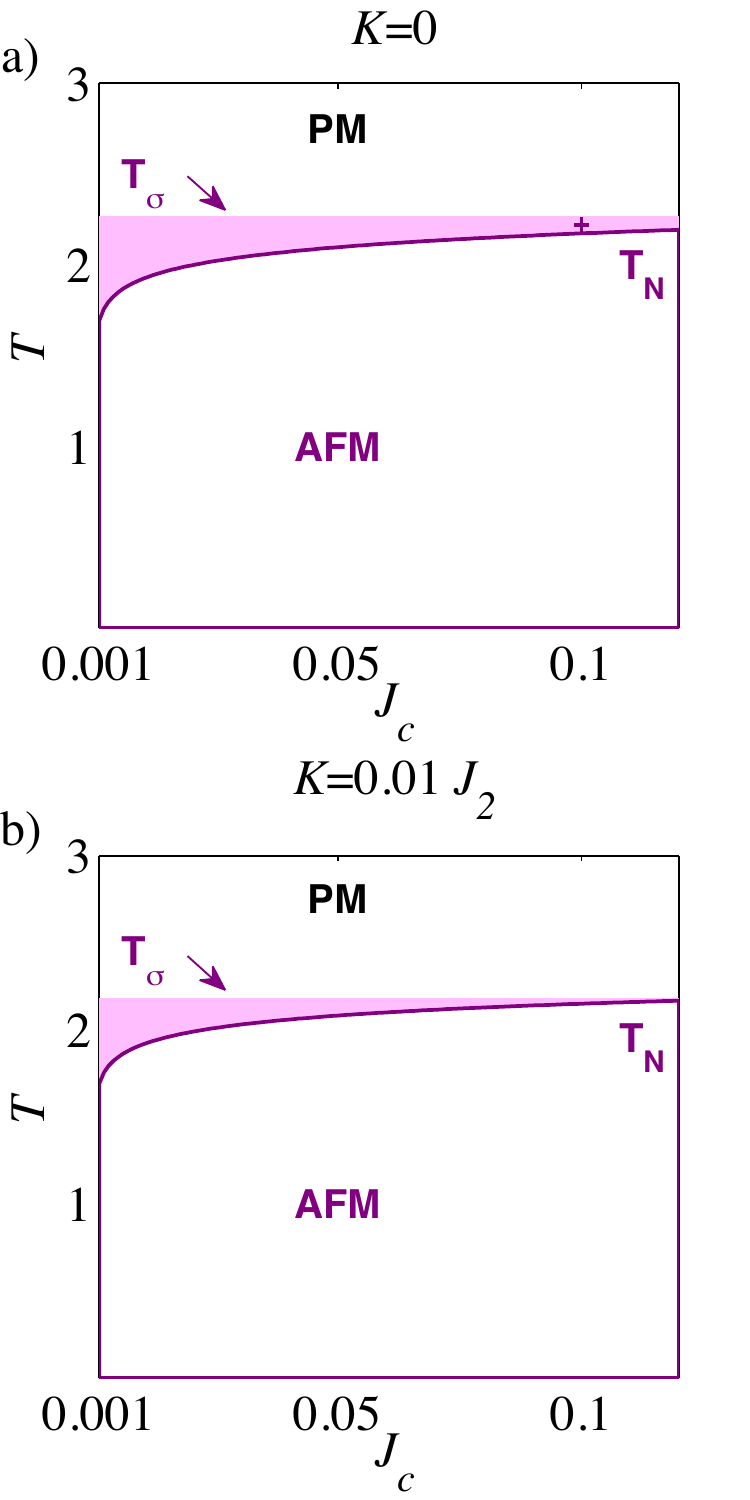}
\caption{(Color online) $T$ vs. $J_c$ phase diagrams for the case of $S=1$ with a biquadratic coupling of a) $K=0$ and b) $K=0.01J_2$. The colored region corresponds to the Ising-nematic regime, which covers an appreciable range of temperatures only for unrealistically small values of $J_c$ (which turn this into an effective two-dimensional problem) and requires very precise fine-tuning for $K$.}
\label{fig:phasediagk}
\end{figure}

Our theoretical calculations based on the effective spin Hamiltonian Eq.~(\ref{eq:J1-J2-K}) confirm the above picture. Namely, we find 
a paramagnetic nematic temperature region $T_N<T<T_\sigma$ where the staggered magnetization vanishes but the correlations along the $x$- and $y$-directions are  different.  
Therefore, we take the quantity 
\beq
\Gamma_y - \Gamma_x \equiv \langle\vec{S}_{\vec{r}}\cdot\vec{S}_{\vec{r}\pm\hat{y}}\rangle -  \langle\vec{S}_{\vec{r}}\cdot\vec{S}_{\vec{r}\pm\hat{x}}\rangle
\label{eq:nematicity}
\eeq
as a measure of electron spin nematicity, which is plotted as a function of temperature in Fig.~\ref{fig:dm_params}.

 The nematic phase is most pronounced in two spatial dimensions ($J_c =0$), where true long-range magnetic order cannot exist by virtue of the Mermin--Wagner theorem~\cite{Mermin1966} whereas the discrete $Z_2$ symmetry can still be broken. However, we find that inclusion of a very small interplanar coupling $J_c$ is sufficient to make $T_N$ approach $T_\sigma$, and the dynamic temperature range $T_\sigma - T_N$ shrinks rapidly as a function of $J_c$, as shown in Fig.~\ref{fig:phasediagk}a). In the absence of a biquadratic spin coupling, this result has already been anticipated in Ref.~\onlinecite{Fang2008} using a large-$N$ approach and in  Ref.~\onlinecite{Goswami2011} using Dyson--Maleev large-S spin-wave theory (see Appendix~\ref{sec:append-DM} for more details of the method). 
 
 Here, we are interested in how the biquadratic spin-spin coupling $-K(\vS_i\cdot \vS_j)^2$ affects the above result. What we found is that the dynamic temperature range of the nematic phase shrinks considerably upon including even a small biquadratic term $K=0.01J_2$, see Fig.~\ref{fig:phasediagk}b). In fact, $T_\sigma$ practically coincides with $T_N$ for sufficiently large $J_c/J_2 > 0.05$.
We conclude that considerable fine-tuning is needed in order to achieve a purely Ising-nematic phase within the model in Eq.~(\ref{eq:J1-J2-K}) and only a very narrow nematic region is observed, which completely disappears for appreciable values of $K$ such as $K\sim 0.6 J_1$ required to fit the INS data~\cite{Wysocki2011}.
This indicates that while the effective spin model~(\ref{eq:J1-J2-K}) is very successful in modeling the spin-wave dispersions (Sec.~\ref{sec:disp}) and INS spin structure factor (Sec.~\ref{sec:struc}), an effectively single orbital spin physics may be insufficient to explain the considerable dynamical range of nematic temperatures observed experimentally in the iron pnictides. Even more striking, the apparent absence of a long-range magnetic order in stoichiometric FeSe~\cite{McQueen2009,Bendele2010} while $T_\sigma\sim90$~K remains large, indicates a 
failure of a pure spin approach and highlights the importance of multi-orbital physics that likely plays an important role in FeSe and in the iron pnictides. This conclusion is corroborated by the recent angle-resolved photoemission spectroscopy (ARPES)~\cite{Shimojima2014} and nuclear magnetic resonance (NMR) studies~\cite{Baek2014,Bohmer2014}  on FeSe.

 \begin{figure*}[!tb]
\centering
\includegraphics[scale=0.6]{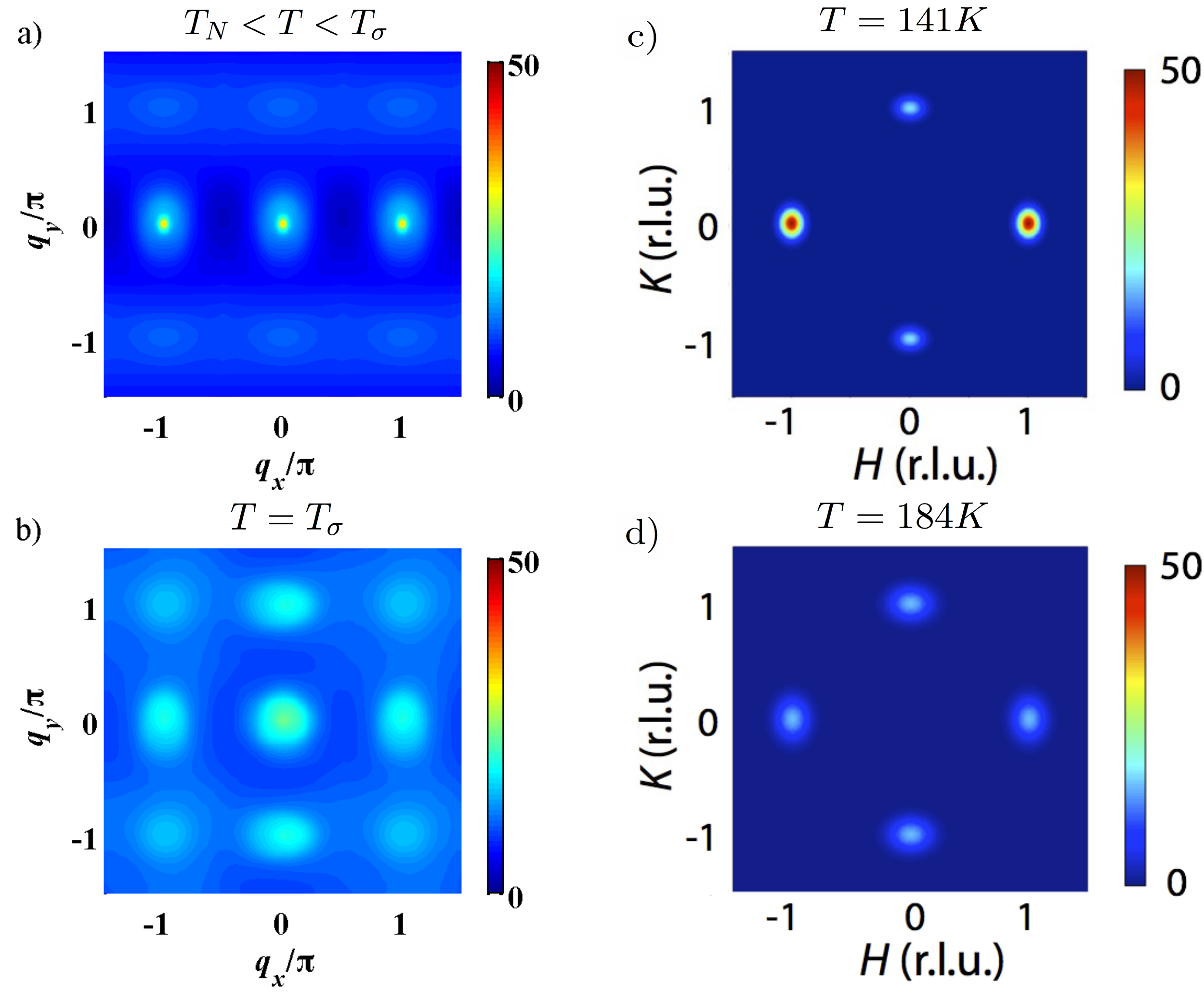}
\caption{(Color online) (a) and (b) Calculated dynamical structure factor $S(\vq,\omega=0.5J_2)$ using the DM method for the case of $S=1$, $K=0$ and $J_c=0.1J_2$  at two temperatures: (a) $T=1.043 T_N$ ($T_N<T<T_{\sigma}$), with its position indicated in the phase diagram of Figs. \ref{fig:phasediagk}a) and \ref{fig:phasediagk}b) $T_\sigma=1.085T_N$. We note that in panel (b), the figure was rotationally symmetrized due to the known limitation of the DM method \cite{Yu2012} which gives an unphysical zero value of $S(\vq,\omega)$ above the temperature $T>T_\sigma$. (c) and (d) Experimental INS data on detwinned BaFe$_2$As$_2$ above the N\'eel temperature $T_N=T_s\approx138$~K, adopted from Lu \emph{et al.} [\onlinecite{Lu2014}].}
\label{fig:dai_exp}
\end{figure*}

However, one must note that while nematic fluctuations have been observed up to a temperature $T^*$ well above $T_N$ \cite{Lu2014}, experiments like this are performed under an applied uniaxial pressure in order to detwin the samples, which explicitly breaks the $C_4$ symmetry. As a result, the observed temperature scale $T^*$ is more indicative of a crossover rather than a true phase transition, and the value of $T^*$ itself becomes likely pressure dependent. Since our theory does not replicate the effect of the applied pressure, it is more sensible to identify the obtained $T_{\sigma}$ with the structural transition temperature $T_s$ instead. If this is the case, then the narrow range of the nematic region found in this work is not surprising, considering that the structural and magnetic transitions are always observed in very close proximity of one another in the Ba-122 family (and actually coincide in the parent compound).

We now proceed to address the issue of recent INS measurements on BaFe$_{2-x}$Ni$_x$As$_2$, which show that the low-energy spin-wave excitations break the $C_4$ symmetry and remain anisotropic even above the structural transition temperature $T_s$, in the nominally tetragonal phase~\cite{Lu2014}.
 We have calculated the dynamical structure factor $S(\vq,\omega)$ in the paramagnetic phase using the Dyson--Maleev spin representation~\cite{Dyson1956,Dyson1956a,Maleev1958} and we indeed find the $C_2$ anisotropy of the intensity in the temperature range $T_N<T<T_\sigma$, similar to the experimental results (see Fig. \ref{fig:dai_exp}). It has to be pointed out, however, that normally, one would associate the temperature $T_\sigma$ with the structural transition temperature $T_s$ because it marks the breaking of the $C_4$ lattice symmetry. Therefore, it is puzzling that the experimentally observed anisotropy persists above $T_s$ [\onlinecite{Lu2014}].
 It has to be remembered, however, that the INS experiment in Ref.~\onlinecite{Lu2014} was done on detwinned crystals, i.e., in the presence of a non-zero uniaxial strain which itself breaks the $C_4$ lattice symmetry. Then, the notion of a spontaneous symmetry breaking no longer applies and strictly speaking, the transition at $T_s$ disappears and instead becomes a crossover. Given the very large nematic susceptibility near $T_s$, as inferred from the resistivity measurements~\cite{Chu2012}, a natural explanation of the neutron scattering data would be that the applied strain triggers a nematic response whose tail is seen at elevated temperatures $T>T_s$. The  spin (and coupled orbital) fluctuations in the nematic channel thus contribute to the observed signal.

\section{\label{sec:disp}Spin-wave dispersions for the magnetic ground-state}

The theory of the Dyson--Maleev (DM) bosons (sometimes also referred to as \emph{modified} spin-wave theory) has already succeeded in giving a good qualitative picture of experimental data in various studies \cite{Goswami2011,Stanek2011,Yu2012}. However, we now put our method of choice to further test by benchmarking its results against those obtained with several other spin-wave theories (see Appendices for more details). For this purpose, we choose to focus on the spin-wave dispersions in the magnetically ordered ground state, and throughout this and the next section, we shall be using the two-dimensional version of our model Hamiltonian, with $J_c=0$. 
In addition to giving us more insight into the validity of the methods typically used for these studies, this analysis serves two other purposes.


First, we note that Stanek \emph{et al.} in Ref. \onlinecite{Stanek2011} have found several discrepancies between the results of the Schwinger boson (SB) representation and those obtained with our method of choice (full decoupling of the DM), with the latter approach producing more accurate results when  a biquadratic spin-spin coupling $K\neq 0$ is included into Eq.~(\ref{eq:J1-J2-K}). Here, we would like to investigate whether this inaccuracy of the SB method pertains to any of the other spin-wave theories, as well as to try to get some insight into its origin. Second, this comparison offers an ideal opportunity to put the recently developed generalized spin-wave theory (GSWT)~\cite{Muniz2013} to the test.

\begin{figure*}[!th]
\centering
\includegraphics[scale=1]{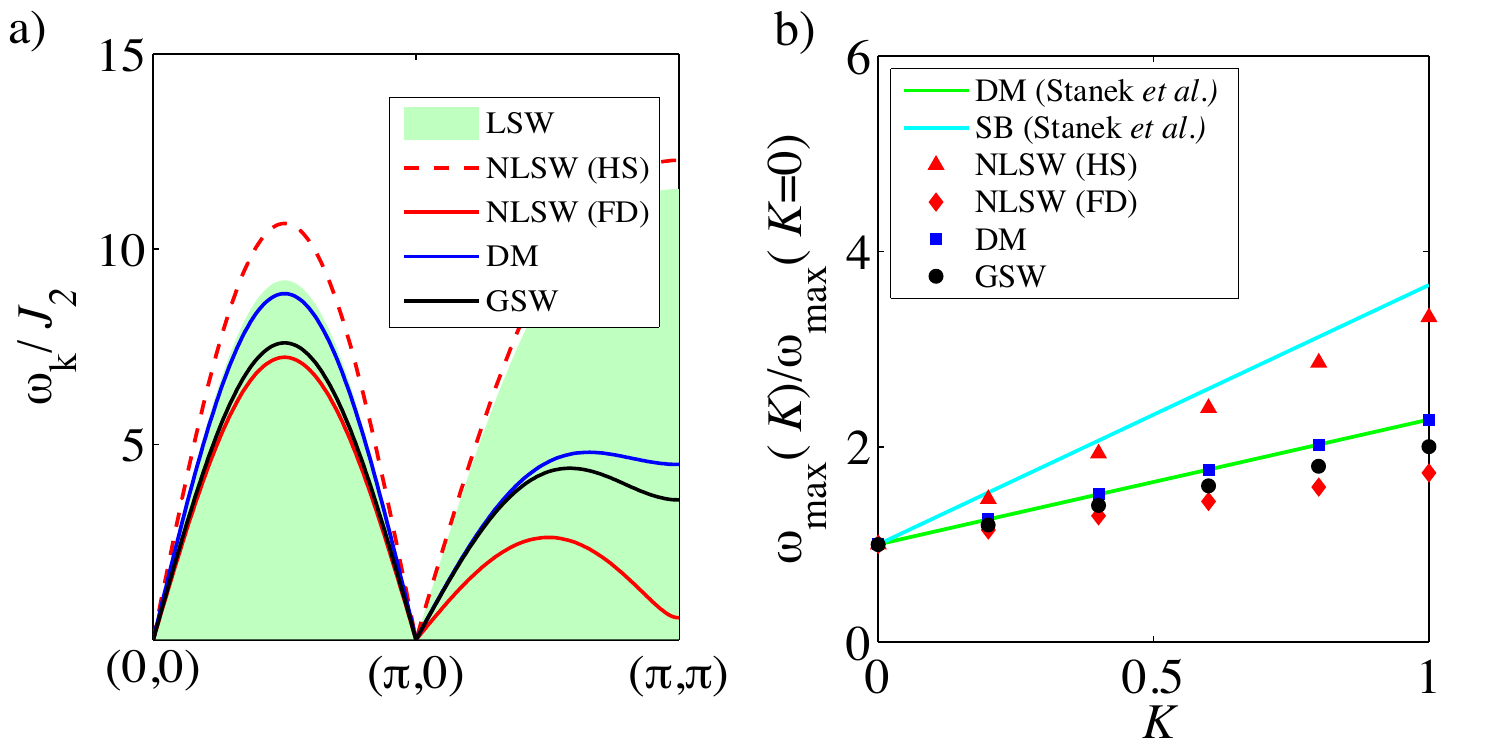}
\caption{(Color online) Discrepancies among the results from the different spin-wave theories. In a) we plot the spin-wave dispersions of the magnetically ordered ground-state of the $J_1-J_2-K$ model for the case of $S=1$, for a value of $K=0.8J_2$, using the results from all the methods we tried. The simplified decoupling of the NLSWT using the HS transformation is almost as ineffective as the LSWT at capturing the effect of the quantum fluctuations whereas using a full decoupling scheme in the NLSWT produces much more accurate results, and agrees favorably with the GSWT and the DM methods.
In b) we plot the evolution of the normalized maximum of the ground-state dispersion [at point $\left(\pi,\pi\right)$] for the $J-K$ model, comparing the results from all our methods to those of Ref. \onlinecite{Stanek2011} (Stanek \emph{et al}). Again, results from the simply decoupled NLSWT overlap with those obtained via SB (known to also underestimate fluctuations) while the rest of the methods account for all the possible correlations to a similar extent.}
\label{fig:s1all}
\end{figure*}

We start with the study of non-linear spin-wave (NLSW) theories, which use a semi-classical approach to the Holstein-Primakoff spin representation by expanding into powers of the small parameter $1/S$ (see Appendix~\ref{sec:append-NLSW} for more details).
There are several different approaches to the $1/S$ expansion. When studying the Heisenberg model, Hamer \emph{et al.}\cite{Hamer1992} used Rayleigh-Schr\"odinger perturbation theory and expanded around the anisotropic (Ising) limit of the Hamiltonian, while Igarashi and Watabe\cite{Igarashi1991} calculated the self-energies to different orders in $1/S$. In both cases, these corrections were made to the diagonalized Bogoliubov dispersion, which only included the lower order (quadratic) terms in the boson creation/annihilation operators. In our case, the presence of the biquadratic spin-spin interaction leads to terms of higher-order in the boson operators, and direct application of the aforementioned methods is highly nontrivial. To proceed, we decouple the higher order terms using Wick's theorem first and then diagonalize the resulting Hamiltonian, with the higher-order terms included on an equal footing into the resulting Bogoliubov dispersion. While this method is conceptually and technically simpler (in particular, it avoids some singular integrals appearing when using the previously mentioned techniques), it results in some differences between our method and those in Refs.~\onlinecite{Igarashi1991,Hamer1992}.
In particular, this procedure does not always guarantee the mandatory existence of the Goldstone modes in the long-range ordered columnar magnetic ground state. It turns out that up to order $\mathcal{O}(S^0)$ in the $1/S$ expansion, the Goldstone modes are correctly reproduced, which is why we limit our expansion to this order. The higher-order corrections in $1/S$, while not captured by this method, are known to be very small for spin $S\geq1$~\cite{Igarashi1991,Hamer1992} and we conclude that our method is therefore sufficiently accurate for our purposes. 
We note that the same approach to non-linear spin-wave theory has been recently used by Stanek~\emph{et al.}\cite{Stanek2011} to analyze the bilinear-biquadratic model for the iron pnictides [see Fig.~\ref{fig:s1all}b) and Table~\ref{table1} for a detailed comparison with their results]. 




 Note that for the purposes of comparing the two approaches, we have performed the Wick decoupling in two different ways. The most straightforward option is to directly decouple all the terms in the Hamiltonian on an equal footing. Throughout the paper we'll refer to this approach as the full decoupling (FD). Alternatively, one can first decouple the biquadratic term via the Hubbard-Stratonovich (HS) transformation in terms of the $\Gamma_{x,y}$ variables defined in Eq.~(\ref{eq:Gamma}), and then decouple the remaining spin bilinears using Wick's theorem for bosons. The latter method results in somewhat simpler expressions for the spin-wave dispersions (compare Eqs.~\ref{eq:AB_HS} and \ref{eq:AB_FD}). However, as we shall show below, the results of the full decoupling scheme are much more accurate.


As a result of the Wick's decoupling, we introduce the following order parameters to be determined self-consistently by the variational principle:
\begin{equation}
\begin{aligned}
&n=\left<a_{\vec{r}}^{\dagger}a_{\vec{r}}\right>\\
&g_x=\left<a_{\vec{r}}a_{\vec{r}+\hat{x}}\right>=\left<a_{\vec{r}}^{\dagger}a_{\vec{r}+\hat{x}}^{\dagger}\right>\\
&f_y=\left<a_{\vec{r}}^{\dagger}a_{\vec{r}+\hat{y}}\right>=\left<a_{\vec{r}}a_{\vec{r}+\hat{y}}^{\dagger}\right>\\
&g_{xy}=\left<a_{\vec{r}}a_{\vec{r}+\hat{x}\pm\hat{y}}\right>=\left<a_{\vec{r}}^{\dagger}a_{\vec{r}+\hat{x}\pm\hat{y}}^{\dagger}\right>
\end{aligned}
\label{eq:parameters}
\end{equation}

As usual, $n$ represents the on-site average number of bosons that decrease the value of the sublattice magnetization $m_s$ from the classical value of $m_\text{class}=S$. The remaining three order parameters correspond to the correlations between neighboring spins in the $\hat{x}$, $\hat{y}$ and $\hat{x}\pm\hat{y}$ directions. It is worth noting that both $g_x$ and $g_{xy}$ are anomalous averages in the sense that the operators involved do not conserve the particle number. This is, however, an artifact of our method of choice, which requires a $\pi$-rotation around, say, $S_x$ direction for one of the sublattices so that the  classical spin orientations now effectively become ferromagnetic. Finally, the averages not included must vanish in order for the total $z$-component of the spin $S_{\text{tot}}^z=\sum_iS_i^z$ to be conserved.


\begin{figure}[!tb]
\centering
\includegraphics[scale=0.75]{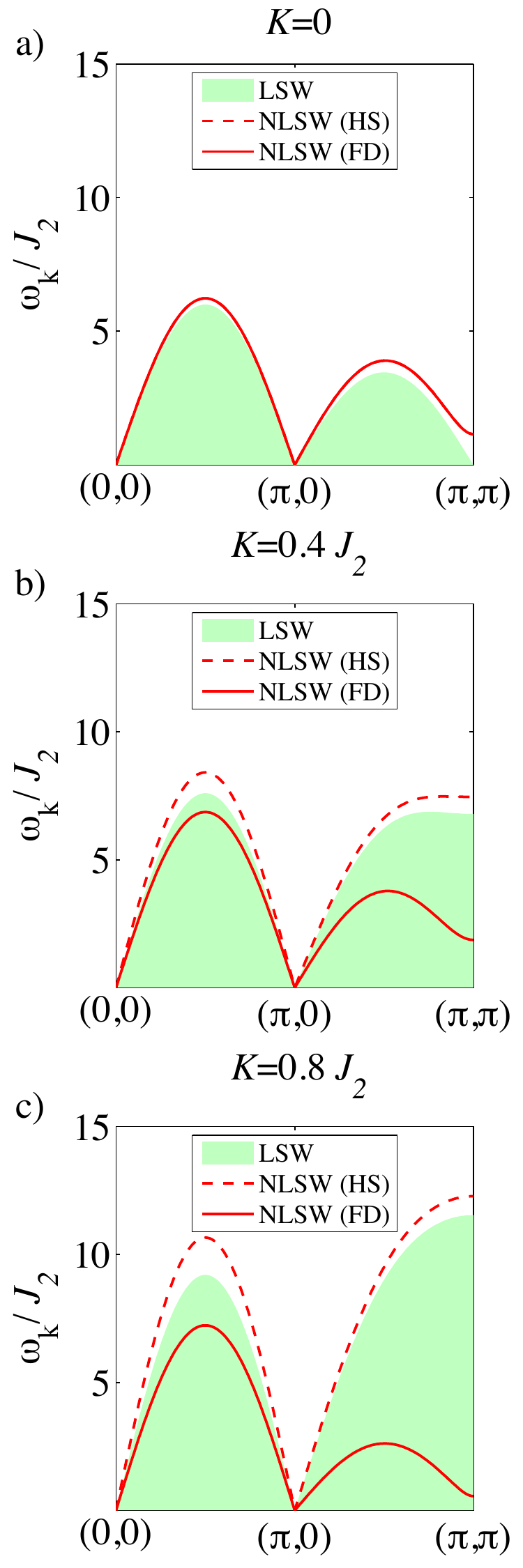}
\caption{(Color online) LSWT and NLSWT results for the spin-wave dispersions of the $S=1$ case for a) $K=0$ b) $K=0.4J_2$ and c) $K=0.8J_2$. In agreement with the results in Ref.~\onlinecite{Stanek2011}, quantum fluctuations suppress the appearance of a maximum at $(\pi,\pi)$ for the fully decoupled NLSWT. In turn, the simpler decoupling scheme based on the Hubbard-Stratonovich transformation with $\Gamma_{x,y}$ bond averages replicates the findings in 
Ref.~\onlinecite{Wysocki2011}, which underestimate the effect of quantum fluctuations.}
\label{fig:s1area}
\end{figure}

\begin{figure}[!tb]
\centering
\includegraphics[scale=0.75]{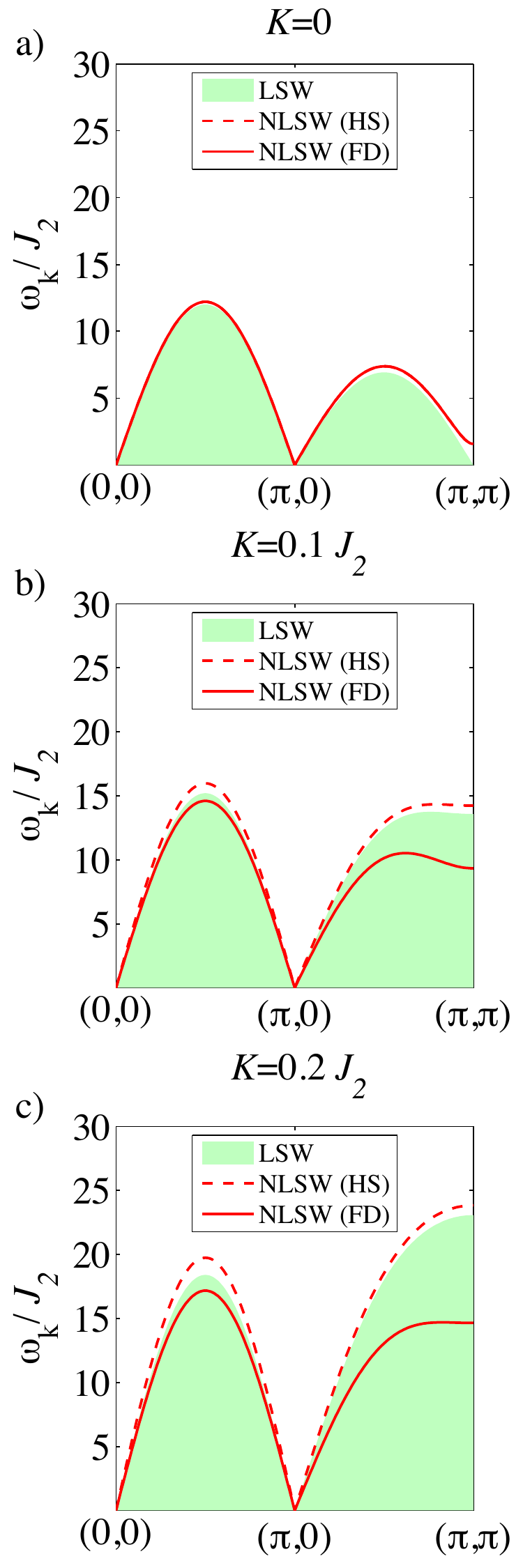}
\caption{(Color online) LSWT and NLSWT results for the spin-wave dispersions of the $S=2$ case for a) $K=0$ b) $K=0.1J_2$ and c) $K=0.2J_2$. The values of the biquadratic constant are reduced by a factor of $1/S^2$ with respect to those of the case of $S=1$ since the constant $K$ is of that order \cite{Wysocki2011}. Note that in this case, the $(\pi,\pi)$ point has become a maximum for all methods, due to the smaller relative importance of the fluctuations for a larger size of spin.}
\label{fig:s2area}
\end{figure}

Once the Hamiltonian is decoupled and diagonalized, we are left with typical Bogoliubov dispersions (see the Appendix~\ref{sec:append-NLSW} for more details). As mentioned earlier, we restrict our analysis up to the order $\mathcal{O}(S^0)$ in the expansion. This should be contrasted with the linear spin-wave (LSW) theory, which corresponds to keeping only the terms of the order $\mathcal{O}(S)$. The spin-wave dispersions calculated in the LSW theory are compared with higher-order non-linear spin-wave (NLSW) theories, see NLSW (HS) and NLSW (FD) in Fig.~\ref{fig:s1area} for $S=1$ and in Fig.~\ref{fig:s2area} for $S=2$. 
Indeed, if we compare the obtained spin-wave dispersions in the magnetic ground state (see Fig.~\ref{fig:s1area}) we observe that, although both orders give similar results for the purely bilinear case ($K=0$), the discrepancies increase dramatically with the value of the biquadratic coupling, especially around the zone boundary or $M=\left(\pi,\pi\right)$ point. More specifically, the region around the $M$ point remains a local minimum of the spin-wave dispersion in all cases for the full decoupling results, while it progressively becomes a local maximum for higher values of $K$ in the Hubbard-Stratonovich approximation. This difference becomes less pronounced as we increase the value of spin (see Fig.~\ref{fig:s2area} for the case of $S=2$), which is readily understood since quantum fluctuations are diminished as $S$ grows and becomes more classical.

For reference, we include the spin-wave dispersions that we obtain from a simple linear spin-wave theory (LSWT) by keeping only the quadratic terms in the Hamiltonian, without performing any decoupling. The simply decoupled non-linear results $(HS)$ almost overlap with those of the LSWT, reinforcing the conclusion that the full decoupling $(FD)$ constitutes a significant improvement over the former method.
We note that quantum fluctuations suppress the appearance of a maximum at $(\pi,\pi)$ for the fully decoupled NLSWT (see Figs. \ref{fig:s1area} and \ref{fig:s2area}), in agreement with the results in Ref. \onlinecite{Stanek2011}. In turn, simply decoupled NLSWT results replicate the findings in Ref. \onlinecite{Wysocki2011}, which underestimate the effect of quantum fluctuations.

Another method used for benchmarking is the recently developed generalized spin-wave theory  (GSWT) \cite{Muniz2013}. This approach is also semi-classical in the sense that it still involves an expansion about a small parameter that measures the deviations from a purely classical ground state. However, it has the advantage of using the fundamental representation of the $SU(N)$ group instead of $SU(2)$ (in our case, $N=2S+1=3$ for $S=1$), designed to capture spin-quadrupolar order in addition to the dipolar magnetic order\cite{Penc-review}.
As we show below, this approach also allows for a more accurate treatment of the biquadratic spin-spin interaction. We note that unlike the usual spin-wave theory, the GSWT introduces several bosonic modes, so we always have $m=2S$ different dispersions instead of a single one. We only plot the lowest-energy mode, since this is the one that describes the low-lying spin-wave excitations.

When plotting the results of all these alternate approaches side-by-side with the DM results [Fig. \ref{fig:s1all}a)], we can readily check that the fully decoupled NLSWT, GSWT and our main choice, the also fully decoupled DM all reflect accurately the effects of quantum fluctuations. In the case of the GSWT, this is achieved by the inclusion of the most general order parameter, largely improving over the usual LSWT, which only accounts for fluctuations around the \emph{classical} vector field. For the DM approach and NLSWT with full decoupling, it is essential to include all the possible spin correlations, which is achieved by performing a full decoupling of the biquadratic term (see Appendix~\ref{sec:append-DM}), rather than the Hubbard-Stratonovich transformation used in Eq.~(\ref{eq:mean-field}). 
It is worth noting that such a full decoupling is possible in the DM approach while still preserving the Goldstone modes because no $1/S$ terms are necessary when using the DM bosons. This is precisely what also makes it our method of choice for the analyses at higher temperatures. 

For completeness, we also apply all of the above techniques to the simpler $J-K$ model with a N\'eel ground state ($J_2=0$), in order to further the comparison between different methods in Ref.~\onlinecite{Stanek2011}. We obtain the maxima of the dispersions at the Brillouin zone edge which are shown in Fig. \ref{fig:s1all}b) for several different values of the biquadratic coupling $K$ (the values are normalized with respect to the case of $K=0$ for ease of comparison).
As expected, the results for the simply decoupled NLSWT overlap with the curve for the SB results. As explained above, the overestimation of the dispersion maxima originates from the inability of these approaches to correctly capture the effect of quantum fluctuations.
By contrast, full decoupling accomplished by the NLSWT and the DM methods, as well as the GSWT all produce slopes of the spin-wave dispersion that are nearly identical to  those found by exact diagonalization in Ref.~\onlinecite{Stanek2011}.
A simple metric that allows one to compare different approaches is the tangent of the slope $\ud\, \omega_\text{max}(K)/\ud K$ of the spin-wave dispersion maxima shown in Fig.~\ref{fig:s1all}b). The numerical values of the slopes obtained with all these different methods are shown in Table \ref{tab:table1}.

We must note that we are using a self-consistent set of Euler--Lagrange equations to solve for the expectation values of the variables in Eq.~(\ref{eq:parameters}), whereas Stanek \emph{et al.}\cite{Stanek2011} directly minimize the free energy in the $J-K$ model case. It turns out that this latter method is much harder to implement in the case of the frustrated magnets such as the $J_1-J_2-K$ model studied here, which is why we stick with the self-consistent Euler--Lagrange approach in this study. Of course the physical results are identical no matter which of these two equivalent techniques one uses, as evidenced by the identical slopes in Table~\ref{table1} and in Fig.~\ref{fig:s1all}b) obtained with the Dyson--Maleev method in this work.


\begin{table}
\caption{\label{tab:table1} Comparison of the spin-wave dispersion slopes $\mathrm{d} \omega_{max}/\mathrm{d} K$ calculated with different methods in this work and in Ref.~\onlinecite{Stanek2011} (Stanek \emph{et al.}). Here $\omega_{max}$ is the maximum of the spin-wave dispersion at $(\pi,\pi)$ and the slope is obtained from a linear fit of the date in Fig. \ref{fig:s1all}b).}
\begin{ruledtabular}
\begin{tabular}{lr}
Method&Slope\\
\hline
DM (Stanek \emph{et al.}) & 1.2771\\
SB (Stanek \emph{et al.}) & 2.6567\\
NLSW (Simple Decoupling) & 2.3284\\
NLSW (Full Decoupling) & 0.7319\\
DM & 1.2771\\
GSW & 1.0000
\end{tabular}
\end{ruledtabular}
\label{table1}
\end{table}

\section{\label{sec:struc}Dynamical structure factors}

\begin{figure}[!tb]
\centering
\includegraphics[scale=0.65]{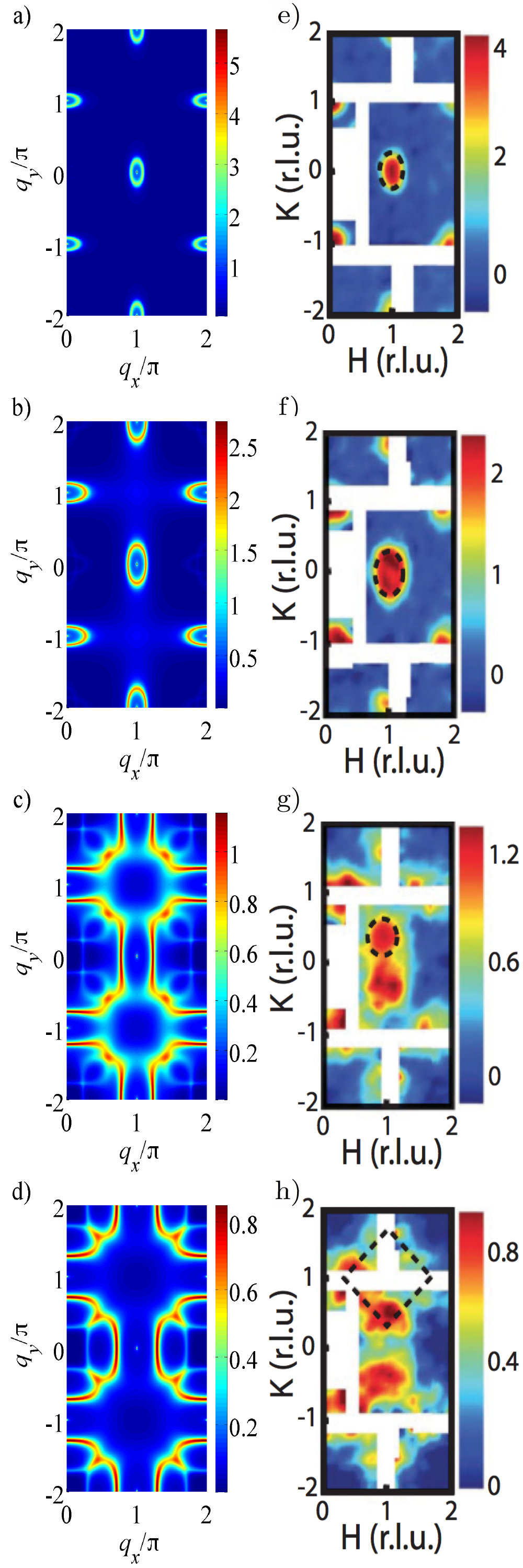}
\caption{(Color online) Comparison of the calculated (left panels) and experimentally measured (right panels) dynamical structure factors at low temperature ($T=7$ K in Ref.~ [\onlinecite{Harriger2011}]). Calculations were performed using GSWT at $K=0.8$ for energy cuts of a) $\omega=2J_2$, b) $\omega=3J_2$, c) $\omega=5J_2$ and d) $\omega=6J_2$. A value of $J_2=25$~meV was used with broadening $\gamma=0.5 J_2$ in Eq.~(\ref{eq:S(T=0)}).
}
\label{fig:struc}
\end{figure}

The differential cross section of the inelastic neutron scattering is proportional to the dynamical spin structure factor $S(\vq,\omega) = \int\ud t\, e^{i\omega t}\mean{\vS_{\vq}(t)\cdot \vS_{-\vq}(0)}$ which measures spin-spin correlations. At zero temperature, only the transverse components of $\mathcal{S}^{xx}(\vec{q},\omega)=\mathcal{S}^{yy}(\vec{q},\omega)$ contribute, which can be expressed as follows \cite{Stanek2011}:
\begin{equation}
\mathcal{S}^{xx}(\vec{q},\omega)=N_s\pi S^{\text{eff}}\frac{\left(\mu_{\vec{q}0}-\Delta_{\vec{q}0}\right)}{\omega_{\vec{q}0}}\delta(\omega-\omega_{\vec{q}0}),
\label{eq:S(T=0)}
\end{equation}
where the coefficients $\mu_{\vq0}$ and $\Delta_{\vq0}$ are given by the expressions in Eqs.~(\ref{eq:mu}) and~(\ref{eq:delta}).

At finite temperatures, longitudinal correlations ($\mathcal{S}^{zz}$) must also be taken into account. This gives a somewhat more complex formula\cite{Yu2012}:
\begin{equation}
\begin{aligned}
\mathcal{S}(\vec{q},\omega)=&\frac{2\pi}{N_s} \sum_{\vec{k}}\sum_{s,s'=\pm 1}\left[\cosh{\left(2\theta_{\vec{k}+\vec{q}}-2\theta_{\vec{k}}\right)}-ss'\right]\\
&\times\delta\left(\omega-s\epsilon_{\vec{k}+\vec{q}}-s'\epsilon-{\vec{k}}\right)n_{\vec{k}+\vec{q}}^{s}n_{\vec{k}}^{s'}
\end{aligned}
\label{eq:S(T)}
\end{equation}
The momentum-dependent angle $\theta_q$ is obtained from the particular Bogoliubov transformation and is given by $\tanh{(2\theta_q)}=\frac{B_{\vec{q}}}{A_{\vec{q}}}$.
We use the notation $n_\vec{k}^{\pm}$ to refer to $n_{\vec{k}}^{-}=n_{\vec{k}}$ and $n_{\vec{k}}^{+}=n_{\vec{k}}+1$, respectively, where $n_\vk$ is the Bose distribution function evaluated at the spin-wave frequency $\omega_k=2\sqrt{|A_\vk|^2 - |B_\vk|^2}$. 

To obtain finite results, we substitute the $\delta$-function in Eqs.~(\ref{eq:S(T=0)}) and (\ref{eq:S(T)}) by a Lorentzian  broadening:
\begin{equation}
\delta(\omega-\Delta\epsilon)\rightarrow\frac{1}{\pi}\frac{\gamma}{(\omega-\Delta\epsilon)^2+\gamma^2}
\end{equation}
The width $\gamma$ includes the instrumental broadening used to mimic the finite experimental resolution and, more importantly, it also incorporates the Landau damping effect due to coupling of spin waves to itinerant electrons. Calculating the magnitude of $\gamma$ would require a detailed microscopic theory that is beyond the scope of this article. Instead, we use $\gamma=0.5 J_2$, somewhat smaller than the value deduced from INS  data on CaFe$_2$As$_2$ [\onlinecite{Goswami2011}] and used in previous theoretical works~\cite{Yu2012}. We find that including larger values of $\gamma \gtrsim J_2$ renders the transverse contribution in Eq.~(\ref{eq:S(T=0)}) nearly featureless, and we find that the main non-trivial effect of broadening is on the longitudinal component in Eq.~(\ref{eq:S(T)}).

We have used both the DM method (see Fig.~\ref{fig:dai_exp}) and the generalized spin-wave theory (see Fig.~\ref{fig:struc}) to compute the dynamical spin structure factor and compared it with the INS experiments on BaFe$_2$As$_2$ from Refs.~\onlinecite{Lu2014} and \onlinecite{Harriger2011}.
At high temperatures, $T>T_N$,  we used the DM method to compare with the recent experiments\cite{Lu2014}, since the DM bosons faithfully capture large deviations from the classical ground state even in the disordered phase~\cite{Yu2012}.
The results of this comparison are plotted in Fig.~\ref{fig:dai_exp} and have already been discussed in Section~\ref{sec:results}. 

To compute the spin structure factor at low temperatures, we choose the GSWT method due to its simplicity, having already checked that its accuracy is comparable to that of the fully decoupled NLSWT and the DM method in the previous section. Indeed, both the DM and GSWT produce nearly identical results when using Eq.~(\ref{eq:S(T=0)}) at $T=0$, but the GSWT is much more straightforward to implement (see Appendix~\ref{sec:append-GSWT} for more details).
To compare with the experimental results, we choose to plot four different energy cuts in Fig.~\ref{fig:struc} and we have rotationally symmetrized the results so that they can be directly compared with the $C_4$-symmetric results observed in the twinned samples of BaFe$_2$As$_2$ in Ref.~\onlinecite{Harriger2011}.
 

The positions of the peaks in $S(\vq,\omega)$ are determined by the condition of energy conservation in the $\delta$-function in Eqs.~(\ref{eq:S(T=0)}) and (\ref{eq:S(T)}), which dictates that at each frequency, maxima will appear at $\omega\approx\omega_{\vec{q}}$. 
For the lower energy cuts in Fig.~\ref{fig:struc}a) and \ref{fig:struc}b), we observe rings that appear centered at $\vec{q_1}=(\pi,0)$ and $\vec{q_2}=(0,\pi)$, the two possible degenerate ordering wave-vectors. The ellipticity of the rings is an indicator of the anisotropy of the system. As we increase the energy of the cuts, the rings expand towards the magnetic zone boundary eventually shifting their peaks from $\vec{q_1}$ and $\vec{q_2}$ to $\vec{q}=\left(\frac{\pi}{2},\frac{\pi}{2}\right)$, as Figs.~\ref{fig:struc}c) and \ref{fig:struc}d) demonstrate. Our results are in semi-qualitative agreement with experimental data~\cite{Harriger2011} (shown in the right panels of Fig.~\ref{fig:struc} for comparison). The main discrepancies are most likely due to a larger effective broadening in the real compounds, produced by the Landau damping as mentioned earlier.


\begin{figure*}[!tb]
\centering
\includegraphics[scale=1.0]{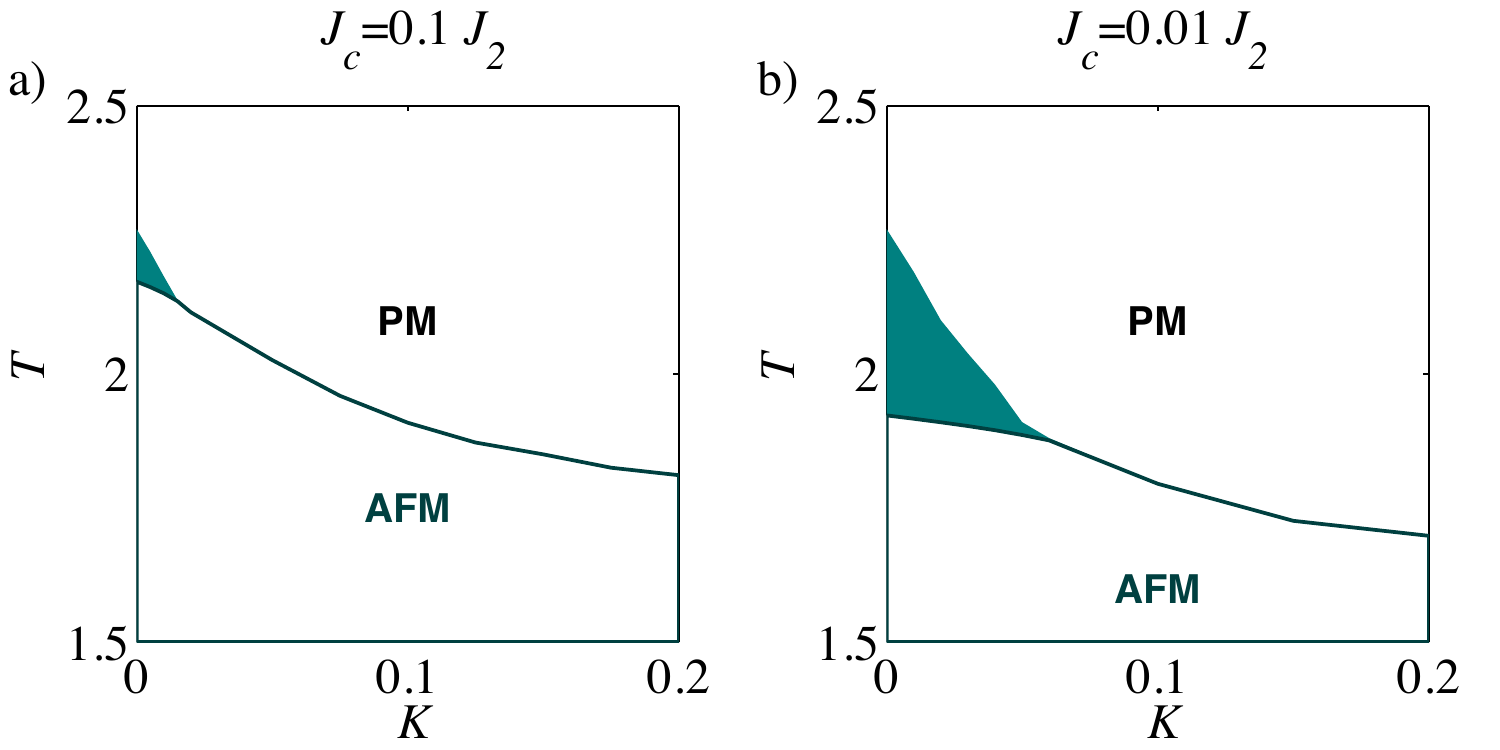}
\caption{(Color online) $T$ vs. $K$ phase diagrams for the case of $S=1$ with values of the interlayer coupling of a) $J_c=0.1J_2$ and b) $J_c=0.01J_2$, respectively. The colored area corresponds to the Ising-nematic region. Both temperatures decrease rapidly with $K$ and coincide fast, with wider nematic regions still appreciable for smaller interlayer couplings.}
\label{fig:phasediagjc}
\end{figure*}

\section{\label{sec:nem}Evolution of nematicity}

In recent neutron scattering experiments \cite{Harriger2011, Ewings2011,Lu2014}, anisotropies have been measured in the iron pnictides, even in the tetragonal, paramagnetic phase above $T_s$. This suggests that the inhomogeneities are not linked to the structural changes, but rather to the magnetic fluctuations. This idea is further supported by the recent nematic susceptibility measurements inferred from electric resistivity in detwinned samples  of BaFe$_2$As$_2$ [\onlinecite{Chu2012}]. Therefore, the structural lattice instability would not be the cause of nematicity. On the contrary, incipient magnetic fluctuations in the nematic regime appear to be responsible for the structural changes.

In two-dimensional spin systems studied previously~\cite{Fang2008,Xu2008,Yu2012}, the Ising-nematic transition is possible even at finite temperature \cite{Chandra1990} since it does not break any continuous symmetry and is not subject to the Mermin--Wagner theorem \cite{Mermin1966}. Here we study the nematic order in three spatial dimensions, by  analyzing the temperature evolution of both the staggered magnetization $m_s$ and the electronic nematicity $(\Gamma_y - \Gamma_x)$ in Eq.~(\ref{eq:nematicity}). 
For reasons already mentioned in section IV, namely the lack of a small expansion parameter, the DM boson representation is our method of choice due to its reliability even deep into the paramagnetic phase. We use the full decoupling of the biquadratic term, considering all channels by means of Wick's theorem as outlined in Ref.~\onlinecite{Holt2011}, and contrast them with the result of the simpler decoupling via Hubbard-Stratonovich [see Eq.~(\ref{eq:mean-field})] used in the earlier work by one of the authors~\cite{Yu2012}.

We define $T_{\sigma}$ as the temperature at which the spin-spin correlations $\Gamma_x$ and $\Gamma_y$ in Eq.~(\ref{eq:nematicity}) become equal in both crystallographic directions so that the system recovers the full $C_4$ symmetry of the tetragonal lattice. As we anticipated in section III, we find that the staggered magnetization $m_s$ always vanishes at a lower or equal temperature than that at which the nematicity does, that is, $T_N\leq T_{\sigma}$. We refer to the region $T_N\leq T\leq T_{\sigma}$ as the pure Ising-nematic phase. However, we find that for typical values of $J_c$ and $K$ (see Fig. \ref{fig:dm_params}) the temperature range of this phase is either very narrow or non-existent. In order to find out the origin of this behavior, we perform an exhaustive analysis of the $T_{\sigma}-T_N$ range as a function of the two tunable parameters $J_c$ and $K$.

The evolutions of both $T_N$ and $T_\sigma$ have been previously studied for the case with no biquadratic term ($K=0$) in Ref. \onlinecite{Goswami2011}. As expected, we obtain the same qualitative behavior [Fig. \ref{fig:phasediagk}a)] as other authors, where $T_{\sigma}$ stays constant with a changing inter-planar coupling, whereas $T_N$ is zero in two spatial dimensions but quickly approaches $T_{\sigma}$ as soon as even a small value is allowed for $J_c$. This is the expected behavior since tuning $J_c$ amounts to varying the dimensionality of the system and thus as soon as we enter the three-dimensional regime the magnetic long-range order stabilizes fast, while $T_{\sigma}$ still varies weakly. 
In this work, we further investigate how these results are affected by the inclusion of a non-zero biquadratic coupling $K$.

What we find is that $K$ has a dramatic effect and leads to the very quick narrowing of the dynamical temperature range $T_\sigma - T_N$ even for small values of $K$ and realistic $J_c$ [Fig. \ref{fig:phasediagk}b)]. 
Similarly, one can fix $J_c$ and study the evolution of the range $T_{\sigma}-T_N$ as function of the biquadratic spin coupling $K$. The results are plotted in Fig.~\ref{fig:phasediagjc}, which shows that $T_N$ quickly approaches $T_\sigma$ as $K$ is increased.
This behavior contrasts with the result in Ref.~\onlinecite{Yu2012} where the authors find an increase of both transition temperatures with $K$. Because a simple  mean-field decoupling as that in Eq.~(\ref{eq:mean-field}) was used in Ref.~\onlinecite{Yu2012}, their results are easily explained in terms of \emph{effective} couplings  $J_1^{\text{eff, x(y)}}=J_1-2K\Gamma_{x(y)}$, where the antiferromagnetic character in the $x$-direction in enhanced due to the larger effective value of the coupling in this direction (note that $\Gamma_x<0$), while the coupling in the $y$ direction becomes smaller. Thus, within this simpler mean-field picture, $K$ is clearly responsible for enhancing the anisotropy, so that the $C_2$ symmetry breaking becomes more stable for larger $K$. In our Dyson--Maleev treatment, we use a more complex full decoupling method~\cite{Holt2011}, which more accurately accounts for quantum fluctuations.
Ironically, this results in the diminished regime of stability of the Ising nematic phase as $K$ increases (see Fig.~\ref{fig:phasediagjc}).

We conclude that the pure nematic regime disappears for realistic values of $J_c$ and $K$, so that considerable fine-tuning of these parameters  is necessary for Ising nematicity to occur in an appreciable temperature range above $T_N$.


\section{\label{sec:concl}Conclusions}

We studied the frustrated bilinear-biquadratic spin model applied to the iron pnictides, both in the magnetically ordered phase where the compounds form columnar antiferromagnets,  as well as above the magnetic transition temperature, in the paramagnetic regime. We found signatures of a pure Ising-nematic phase in the form of a $C_2$ anisotropy in the spin-spin dynamical structure factors, one which persists even for $T>T_N$. We identify the temperature ($T_{\sigma}$) at which the system recovers the full $C_4$ symmetry of the spin response  with the physical structural transition temperature $T_s$.

However, recent experiments on Ni-doped BaFe$_2$As$_2$ have found  anisotropies in the spin excitations even beyond this point\cite{Lu2014}, up to some temperature $T^{\ast} > T_s$. The lack of ability of our theory to capture these anisotropic features above $T_{\sigma}$ can be explained in one of the two ways. The first scenario, is that the spin response anisotropy is not static but rather originates from nematic fluctuations alone. Since we use static order parameters for the decoupling of the Hamiltonian in the spin-wave theories [see e.g. Eq.~(\ref{eq:parameters})], this may be the reason why these features are not captured in our theoretical results above $T_{\sigma}$. The second scenario is that in the absence of applied uniaxial strain, the $C_4$ symmetry is indeed restored above the temperature $T_s$ which we identify with $T_\sigma$ in our theory. However, the effect of uniaxial strain used to detwin the samples is to smear the transition, making it a smooth crossover with $C_2$ anisotropies that persist up to some higher temperature $T^{\ast}$. 
Given the very large nematic susceptibility near $T_s$, as inferred from the resistivity measurements~\cite{Chu2012}, this would be a natural explanation since the applied strain would be expected to trigger a nematic response whose tail is seen at elevated temperatures $T>T_s$.
If this is indeed the case, one would expect the crossover temperature  $T^{\ast}$ to be strain-dependent. Future experiments under variable strain would therefore be very desirable to help clarify this issue.

The present study improves upon previous work by other authors~\cite{Goswami2011, Yu2012} by including the effect of the biquadratic coupling $-K(\vS_i\cdot \vS_j)^2$ to study  the evolution of the dynamical temperature range $T_{\sigma}-T_N$ of the Ising-nematic phase. We confirmed that this range decreases rapidly with the inclusion of even a small finite interlayer coupling $J_c$, since it stabilizes the magnetic long range order. Furthermore, we found that the inclusion of $K$ has a similar effect, requiring precise fine-tuning in order to get a pure nematic phase over an appreciable temperature range. 
Unlike the previous work in Ref.~\onlinecite{Yu2012} which adopted a simple Hubbard-Stratonovich treatment  of the biquadratic term [see Eq.~(\ref{eq:mean-field})], we have
used a more accurate scheme that accounts better for fluctuations by using a variety of theoretical techniques (non-linear spin-wave theory, Dyson--Maleev method, and GSWT). In all cases, we found that the present approach results in a  drastically different spin-wave dispersion near the Brilloin zone boundary compared to  that from Ref.~\onlinecite{Yu2012}. 
Similarly, the evolution of the transition temperatures ($T_N$ and $T_{\sigma}$) with increasing $K$ is also very different in the present work, reflecting the higher accuracy of the full decoupling that we employed.


On a more technical level, we have benchmarked several spin-wave theories by comparing the spin-wave dispersions and the dynamical spin structure factors to the inelastic neutron scattering experiments. 
We found our main method of choice to be the full decoupling of the Dyson--Maleev modified spin-wave theory, which accurately captures  the effect of quantum and thermal fluctuations and produces a good semi-quantitative agreement with the INS experiment, especially at higher temperatures $T\gtrsim T_N$.  
 Finally, the recently developed generalized spin wave theory (GSWT)~\cite{Muniz2013} deserves a special mention due to its elegance and simplicity, while providing results comparable in accuracy to those obtained with the fully decoupled DM and NLSWT methods. However, the GSWT fails closer to $T_N$ when the thermal fluctuations significantly reduce the ordered moment.

\begin{acknowledgments}
The authors are grateful to Pengcheng Dai for providing experimental Figs.~\ref{fig:dai_exp}c), d) and to Zhentao Wang and Yang-Zhi Chou for helpful discussion.
 The authors acknowledge financial support from the NSF CAREER award no. DMR-1350237 and the Cottrell Scholar Award from Research Corporation for Science Advancement (grant no. 22799). PBE was also supported by the Robert A. Welch Foundation grant C-1818. 
\end{acknowledgments}

\appendix

\section{Non-linear spin-wave theory}
\label{sec:append-NLSW}

As mentioned before, our starting point is the frustrated Heisenberg Hamiltonian with additional biquadratic coupling.

\begin{equation}
\begin{aligned}
\mathcal{H}_{2D}=&J_1\sum_{<\vec{r},\vec{r'}>}\vec{S}_{\vec{r}}\cdot\vec{S}_{\vec{r'}}+J_2\sum_{<<\vec{r},\vec{r'}>>}\vec{S}_{\vec{r}}\cdot\vec{S}_{\vec{r'}}-\\
&-K\sum_{<\vec{r},\vec{r'}>}\left(\vec{S}_{\vec{r}}\cdot\vec{S}_{\vec{r'}}\right)^2
\end{aligned}
\end{equation}

We will concentrate on the regime of parameters $J_2/J_1>1/2$ (particularly, we choose $J_1=J_2$ throughout the entire paper) so that the lattice is in the columnar AFM phase with ordering wave-vectors $\vec{Q}=(\pi,0)$ or $\vec{Q}=(0,\pi)$, evidenced by neutron scattering experiments. We can then consider two interpenetrating sublattices $A$ and $B$, and sum over all points $\vec{r}\in A$ and $\vec{r'}\in B$. Since the spins in sublattice $B$ are aligned antiferromagnetically with respect to those in sublattice $A$, we perform a rotation by $\pi$ in the former, sending $S_{\vec{r'}}^x\rightarrow\widetilde{S}_{\vec{r'}}^x=S_{\vec{r'}}^x$, $S_{\vec{r'}}^y\rightarrow\widetilde{S}_{\vec{r'}}^y=-S_{\vec{r'}}^y$, and $S_{\vec{r'}}^z\rightarrow\widetilde{S}_{\vec{r'}}^z=-S_{\vec{r'}}^z$.


In our first approach, we write the spin operators via the well established Holstein-Primakoff representation, in terms of the bosonic operators $a_{\vec{r}}^{\dagger}$, $a_{\vec{r}}$:

\begin{equation}
\begin{aligned}
S^+_{\vec{r}}=&\left(\sqrt{2S-a_{\vec{r}}^{\dagger}a_{\vec{r}}}\right)a_{\vec{r}}\\
S^-_{\vec{r}}=&a_{\vec{r}}^{\dagger}\left(\sqrt{2S-a_{\vec{r}}^{\dagger}a_{\vec{r}}}\right)\\
S^z_{\vec{r}}=&S-a_{\vec{r}}^{\dagger}a_{\vec{r}}
\end{aligned}
\end{equation}

After performing the rotation of the $B$ sublattice, the Heisenberg terms in the Hamiltonian take the following form, for the antiferromagnetically and ferromagnetically aligned directions, respectively.

\begin{equation}
\begin{aligned}
&\left(\vec{S}_{\vec{r}}\cdot\vec{S}_{\vec{r'}}\right)_{AFM}=-S^2+S\left(a_{\vec{r}}^{\dagger}a_{\vec{r}}+a_{\vec{r'}}^{\dagger}a_{\vec{r'}}+\right.\\
&\left.+a_{\vec{r}}a_{\vec{r'}}+a_{\vec{r}}^{\dagger}a_{\vec{r'}}^{\dagger}\right)-a_{\vec{r}}^{\dagger}a_{\vec{r'}}^{\dagger}a_{\vec{r}}a_{\vec{r'}}-\frac{1}{4}\left(a_{\vec{r}}^{\dagger}a_{\vec{r}}a_{\vec{r}}a_{\vec{r'}}+\right.\\
&\left.+a_{\vec{r}}^{\dagger}a_{\vec{r}}^{\dagger}a_{\vec{r'}}^{\dagger}a_{\vec{r}}+a_{\vec{r'}}^{\dagger}a_{\vec{r}}a_{\vec{r'}}a_{\vec{r'}}+a_{\vec{r}}^{\dagger}a_{\vec{r'}}^{\dagger}a_{\vec{r'}}^{\dagger}a_{\vec{r'}}\right)
\end{aligned}
\end{equation}

\begin{equation}
\begin{aligned}
&\left(\vec{S}_{\vec{r}}\cdot\vec{S}_{\vec{r'}}\right)_{FM}=S^2-\left[a_{\vec{r}}^{\dagger}a_{\vec{r}}+a_{\vec{r'}}^{\dagger}a_{\vec{r'}}-\right.\\
&\left.-\left(a_{\vec{r}}^{\dagger}a_{\vec{r'}}+a_{\vec{r'}}^{\dagger}a_{\vec{r}}\right)\right]+a_{\vec{r}}^{\dagger}a_{\vec{r'}}^{\dagger}a_{\vec{r}}a_{\vec{r'}}-\frac{1}{4}\left(a_{\vec{r}}^{\dagger}a_{\vec{r}}^{\dagger}a_{\vec{r}}a_{\vec{r'}}+\right.\\
&\left.+a_{\vec{r}}^{\dagger}a_{\vec{r'}}^{\dagger}a_{\vec{r}}a_{\vec{r}}+a_{\vec{r}}^{\dagger}a_{\vec{r'}}^{\dagger}a_{\vec{r'}}a_{\vec{r'}}+a_{\vec{r'}}^{\dagger}a_{\vec{r'}}^{\dagger}a_{\vec{r}}a_{\vec{r'}}\right)
\end{aligned}
\end{equation}

In turn, the biquadratic terms are given by the following expressions:

\begin{equation}
\begin{aligned}
&\left[\left(\vec{S}_{\vec{r}}\cdot\vec{S}_{\vec{r'}}\right)^2\right]_{AFM}=S^4-2S^2\left(S-1\right)\left(a_{\vec{r}}^{\dagger}a_{\vec{r}}+a_{\vec{r'}}^{\dagger}a_{\vec{r'}}+\right.\\
&\left.+a_{\vec{r}}a_{\vec{r'}}+a_{\vec{r}}^{\dagger}a_{\vec{r'}}^{\dagger}\right)+S^2\left(1+6a_{\vec{r}}^{\dagger}a_{\vec{r'}}^{\dagger}a_{\vec{r}}a_{\vec{r'}}\right)+S^2\left[\right.\\
&\left.\frac{5}{2}\left(a_{\vec{r}}^{\dagger}a_{\vec{r}}a_{\vec{r}}a_{\vec{r'}}+a_{\vec{r}}^{\dagger}a_{\vec{r}}^{\dagger}a_{\vec{r'}}^{\dagger}a_{\vec{r}}+a_{\vec{r'}}^{\dagger}a_{\vec{r}}a_{\vec{r'}}a_{\vec{r'}}+a_{\vec{r}}^{\dagger}a_{\vec{r'}}^{\dagger}a_{\vec{r'}}^{\dagger}a_{\vec{r'}}\right)+\right.\\
&\left.+\left(a_{\vec{r}}^{\dagger}a_{\vec{r}}^{\dagger}a_{\vec{r}}a_{\vec{r}}+a_{\vec{r'}}^{\dagger}a_{\vec{r'}}^{\dagger}a_{\vec{r'}}a_{\vec{r'}}+a_{\vec{r}}a_{\vec{r}}a_{\vec{r'}}a_{\vec{r'}}+a_{\vec{r}}^{\dagger}a_{\vec{r}}^{\dagger}a_{\vec{r'}}^{\dagger}a_{\vec{r'}}^{\dagger}\right)\right]
\end{aligned}
\end{equation}

\begin{equation}
\begin{aligned}
&\left[\left[(\vec{S}_{\vec{r}}\cdot\vec{S}_{\vec{r'}}\right)^2\right]_{FM}=S^4-2S^2\left(S-1\right)\left[a_{\vec{r}}^{\dagger}a_{\vec{r}}+a_{\vec{r'}}^{\dagger}a_{\vec{r'}}-\right.\\
&\left.-\left(a_{\vec{r}}^{\dagger}a_{\vec{r'}}+a_{\vec{r'}}^{\dagger}a_{\vec{r}}\right)\right]+S^2\left(1+6a_{\vec{r}}^{\dagger}a_{\vec{r'}}^{\dagger}a_{\vec{r}}a_{\vec{r'}}\right)-S^2\left[\right.\\
&\left.\frac{5}{2}\left(a_{\vec{r}}^{\dagger}a_{\vec{r}}^{\dagger}a_{\vec{r}}a_{\vec{r'}}+a_{\vec{r}}^{\dagger}a_{\vec{r'}}^{\dagger}a_{\vec{r}}a_{\vec{r}}+a_{\vec{r}}^{\dagger}a_{\vec{r'}}^{\dagger}a_{\vec{r'}}a_{\vec{r'}}+a_{\vec{r'}}^{\dagger}a_{\vec{r'}}^{\dagger}a_{\vec{r}}a_{\vec{r'}}\right)-\right.\\
&\left.-\left(a_{\vec{r}}^{\dagger}a_{\vec{r}}^{\dagger}a_{\vec{r}}a_{\vec{r}}+a_{\vec{r'}}^{\dagger}a_{\vec{r'}}^{\dagger}a_{\vec{r'}}a_{\vec{r'}}+a_{\vec{r}}^{\dagger}a_{\vec{r}}^{\dagger}a_{\vec{r'}}a_{\vec{r'}}+a_{\vec{r'}}^{\dagger}a_{\vec{r'}}^{\dagger}a_{\vec{r}}a_{\vec{r}}\right)\right]
\end{aligned}
\end{equation}

Where we expanded the square roots by taking $a_{\vec{r}}^{\dagger}a_{\vec{r}}/2S$ as our small parameter. For orders of $\mathcal{O}(S^0)$ and higher in $1/S$, this yields terms with more than 2 bosonic operators. To make them solvable, we decouple them by using Wick's theorem and consider all the possible decouplings. We take all averages to be real for convenience, without loss of generality.

\begin{equation}
\begin{aligned}
&n=\left<a_{\vec{r}}^{\dagger}a_{\vec{r}}\right>\\
&g_x=\left<a_{\vec{r}}a_{\vec{r}+\hat{x}}\right>=\left<a_{\vec{r}}^{\dagger}a_{\vec{r}+\hat{x}}^{\dagger}\right>\\
&f_y=\left<a_{\vec{r}}^{\dagger}a_{\vec{r}+\hat{y}}\right>=\left<a_{\vec{r}}a_{\vec{r}+\hat{y}}^{\dagger}\right>\\
&g_{xy}=\left<a_{\vec{r}}a_{\vec{r}+\hat{x}\pm\hat{y}}\right>=\left<a_{\vec{r}}^{\dagger}a_{\vec{r}+\hat{x}\pm\hat{y}}^{\dagger}\right>
\end{aligned}
\label{eq:MFparams}
\end{equation}

We assume the rest of the averages to be zero by virtue of the conservation of the total z-component of the spin ($S^z=\sum_i S^z_i$) in each direction. In principle, both bilinear and biquadratic terms can be treated in this manner. However, for the purpose of studying the differences between the different approaches to the decoupling procedure, we also use a Hubbard-Stratonovich (HS) transformation for the biquadratic spin term:
\begin{equation}
\left(\vec{S}_{\vec{r}}\cdot\vec{S}_{\vec{r'}}\right)^2\simeq 2\left<\vec{S}_{\vec{r}}\cdot\vec{S}_{\vec{r'}}\right>\vec{S}_{\vec{r}}\cdot\vec{S}_{\vec{r'}}-\left<\vec{S}_{\vec{r}}\cdot\vec{S}_{\vec{r'}}\right>^2.
\label{eq:mean-field}
\end{equation}

The remaining spin bilinears are then decoupled as usual per Wick's theorem, whereas the HS averages themselves can be expressed in terms of the mean-field parameters in Eq.~(\ref{eq:MFparams}) as follows:

\begin{equation}
\begin{aligned}
&\Gamma_x=\left<\vec{S}_{\vec{r}}\cdot\vec{S}_{\vec{r}+\hat{x}}\right>=-\left(S-n-g_x\right)^2\\
&\Gamma_y=\left<\vec{S}_{\vec{r}}\cdot\vec{S}_{\vec{r}+\hat{y}}\right>=\left(S-n+f_y\right)^2.
\end{aligned}
\end{equation}

The advantage of using the above HS transformation is that it results in simpler expressions for the spin-wave dispersions (see Eq.~\ref{eq:AB_HS} below). However, as we show in the main text, this comes at a price that the HS decoupling is much worse at capturing the spin fluctuations compared to the full decoupling (FD) method. With this proviso, we show the details of both methods below, but the reader is advised to use the FD method for accurate results.


After full use of Wick's theorem, the non-linear spin-wave theory results in the following quadratic Hamiltonian (up to inessential constant terms):
\begin{equation}
\begin{aligned}
\mathcal{H}_{NLSW}=&\sum_{\vec{k}}\left[A_{\vec{k}}\left(a_{\vec{k}}^{\dagger}a_{\vec{k}}+a_{\vec{-k}}a_{\vec{-k}}^{\dagger}\right)+\right.\\
&\left.+B_{\vec{k}}\left(a_{\vec{k}}a_{\vec{-k}}+a_{\vec{k}}^{\dagger}a_{\vec{-k}}^{\dagger}\right)\right],
\end{aligned}
\end{equation}
which, after the Bogoliubov transformation, is expressed in terms of new boson operators:
\beq
\mathcal{H}_{NLSW}=\sum_{\vec{k}}\omega_{\vec{k}}\left(\alpha_{\vec{k}}^{\dagger}\alpha_{\vec{k}}+\frac{1}{2}\right)
\eeq
with the spin-wave dispersion:
\begin{equation}
\omega_{\vec{k}}=2\sqrt{A_{\vec{k}}^2-B_{\vec{k}}^2},
\end{equation}
where the expressions for the coefficients $A_{\vec{k}}$ and $B_{\vec{k}}$ are given in Eq.~(\ref{eq:AB_HS}) for the HS decoupling and in Eq.~(\ref{eq:AB_FD}) for the full decoupling (FD):
\begin{equation}
\begin{aligned}
A_{\vec{k}(HS)}=&(J_1-2K\Gamma_x)(S-n-g_x)+\\
+&(J_1-2K\Gamma_y)(S-n+f_y)(\cos{k_y}-1)+\\
&+2J_2(S-n-g_{xy});\\
B_{\vec{k}(HS)}=&(J_1-2K\Gamma_x)(S-n-g_x)\cos{k_x}+\\
&+2J_2(S-n-g_{xy})\cos{k_x}\cos{k_y}.
\end{aligned}
\label{eq:AB_HS}
\end{equation}


\begin{equation}
\begin{aligned}
A_{\vec{k}(FD)}=&J_1(S-n-g_x)+\\
+&J_1(S-n+f_y)(\cos{k_y}-1)+\\
&+2J_2(S-n-g_{xy})-\\
&-KS^2\{-S+2[1+5(n+g_x)]\}-\\
&-KS^2\{-S+2[1+5(n-f_y)]\}(1-\cos{k_y});\\
B_{\vec{k}(FD)}=&J_1(S-n-g_x)\cos{k_x}+\\
&+2J_2(S-n-g_{xy})\cos{k_x}\cos{k_y}-\\
&-KS^2\{-S+2[1+5(n+g_x)]\}\cos{k_x}.
\end{aligned}
\label{eq:AB_FD}
\end{equation}





Minimizing the free energy with respect to the mean-field parameters defined in Eq.~(\ref{eq:MFparams}), we finally arrive at a system of Euler--Lagrange equations:
\begin{equation}
\begin{aligned}
\alpha_x=&\frac{1}{N_s}\sum_{\vec{k}}\left(\left<a_{\vec{k}}^{\dagger}a_{\vec{k}}\right>+\left<a_{\vec{k}}a_{\vec{-k}}\right>\cos{k_x}\right)=\\
&=-\frac{1}{2}+\frac{1}{N_s}\sum_{\vec{k}}\left(1+2n_{\vec{k}}\right)\frac{A_{\vec{k}}-B_{\vec{k}}\cos{k_x}}{\omega_{\vec{k}}};\\
\beta_y=&\frac{1}{N_s}\sum_{\vec{k}}\left<a_{\vec{k}}^{\dagger}a_{\vec{k}}\right>\left(1-\cos{k_y}\right)=\\
&=-\frac{1}{2}+\frac{1}{N_s}\sum_{\vec{k}}\left(1+2n_{\vec{k}}\right)\frac{A_{\vec{k}}\left(1-\cos{k_y}\right)}{\omega_{\vec{k}}};\\
\alpha_{xy}=&\frac{1}{N_s}\sum_{\vec{k}}\left(\left<a_{\vec{k}}^{\dagger}a_{\vec{k}}\right>+\left<a_{\vec{k}}a_{\vec{-k}}\right>\cos{k_x}\cos{k_y}\right)=\\
&=-\frac{1}{2}+\frac{1}{N_s}\sum_{\vec{k}}\left(1+2n_{\vec{k}}\right)\frac{A_{\vec{k}}-B_{\vec{k}}\cos{k_x}\cos{k_y}}{\omega_{\vec{k}}},
\label{eq:system}
\end{aligned}
\end{equation}
where for convenience, we have denoted: $\alpha_x=n+g_x$, $\alpha_{xy} = n + g_{xy}$, and $\beta_y=n-f_y$ following the convention in Ref.~\onlinecite{Muniz2013}. The above equations are to be solved self-consistently because their right-hand side depends on the mean-field parameters themselves via Eqs.~(\ref{eq:AB_HS}) and (\ref{eq:AB_FD}).

\section{Generalized spin-wave theory}
\label{sec:append-GSWT}

This approach, proposed recently in Ref.~\onlinecite{Muniz2013} is derived starting from an $SU(N)$ group representation of $N$ Schwinger bosons, instead of the usual two bosons in the case of $SU(2)$ spins. While the usual linear spin-wave theory (LSWT) describes only fluctuations around the classical vector field, the advantage of introducing the $SU(N)$ order parameter is that it describes correlations characterizing not only the spin-dipolar order, but also the multipolar orders. We note in passing that the GSWT is similar in spirit to the so-called flavour-wave theory that was originally introduced to describe spin-1 systems~\cite{Papanicolaou84, Papanicolaou88} and is commonly used to describe spin-multipolar orders, referred to collectively as ``spin-nematic"~\cite{Penc-review} (not to confuse with the term ``Ising-nematic" which in the case of iron-pnictides refers simply to the spatial anisotropy of the bond spin correlations $\Gamma_x - \Gamma_y$).

In the GSWT, the bosons satisfy the local $SU(N)$ constraint on the local number of bosons:
\begin{equation}
\sum_{m=0}^{N-1}b_{\vec{r}m}^{\dagger}b_{\vec{r}m}=N\mathcal{S}
\end{equation}
In the fundamental representation of $SU(N)$ where $N\mathcal{S}=1$, all spin operators can be expressed as a bilinear combination in the bosonic modes. In particular, for the local spin operators where $N=2S+1$, the matrix elements are given by the following expressions.
\begin{equation}
\begin{aligned}
\mathcal{S}_{mm'}^x=&\delta_{m(m'-1)}\frac{\sqrt{(m+1)(2S-m)}}{2}+\\
&+\delta_{(m+1)m'}\frac{\sqrt{(m'+1)(2S-m')}}{2}\\
\mathcal{S}_{mm'}^y=&\delta_{m(m'-1)}\frac{\sqrt{(m+1)(2S-m)}}{2i}-\\
&-\delta_{(m+1)m'}\frac{\sqrt{(m'+1)(2S-m')}}{2i}\\
\mathcal{S}_{mm'}^z=&\delta_{mm'}(S-m)
\end{aligned}
\end{equation}
And the elements of the matrix associated with the bilinear operator (for details, see \cite{Muniz2013}) $S_r^{\nu\mu}S_r^{\nu\mu}$ are:
\begin{equation}
\mathcal{S}_{mm'}^{\nu\mu}=\sum_{m"}\mathcal{S}_{mm"}^{\nu}\mathcal{S}_{m"m'}^{\mu}
\end{equation}

Finally, following the spirit of the Holstein-Primakoff representation, we impose the constraint (6) by requiring that the condensed fraction satisfies:
\begin{equation}
b_{\vec{r}0}^{\dagger}=b_{\vec{r}0}=\sqrt{1-\sum_{m=1}^{N-1}b_{\vec{r}m}^{\dagger}b_{\vec{r}m}}\simeq 1-\frac{1}{2}\sum_{m=1}^{N-1}b_{\vec{r}m}^{\dagger}b_{\vec{r}m}
\end{equation}
Keeping only the quadratic terms will result in an effective Hamiltonian, which we can diagonalize, obtaining $(N-1)$ Bogoliubov dispersions, one for each bosonic mode. Using these representations for the spin operators, we obtain the following hamiltonian (again, up to constant terms):

\begin{equation}
\begin{aligned}
\mathcal{H}_{GSW}=&\sum_{\vec{k},m}\mu_{\vec{k}m}\left(b_{\vec{k}m}^{\dagger}b_{\vec{k}m}+b_{\vec{-k}m}^{\dagger}b_{\vec{-k}m}\right)+\\
&+\Delta_{\vec{k}m}\left(b_{\vec{k}m}^{\dagger}b_{\vec{-k}m}^{\dagger}+b_{\vec{k}m}b_{\vec{-k}m}\right),
\end{aligned}
\end{equation}
with coefficients (in the two-dimensional case) given by:

\begin{equation}
\begin{aligned}
\mu_{\vec{k}1}=&S\left[2J_2+K(2S-1)^2\right]+\\
&+S\left[J_1-2KS(S-1)\right]\cos{k_y}\\
\mu_{\vec{k}2}=&4S\left[J_2+K(S-1)(2S-1)\right]-\\
&-KS(2S-1)\cos{k_y}\\
\mu_{\vec{k}m}=&mS\left[2J_2-K(m-2S)(2S-1)\right]
\label{eq:mu}
\end{aligned}
\end{equation}

\begin{equation}
\begin{aligned}
\Delta_{\vec{k}1}=&\left\{J_1S+KS\left[1+2S(S-1)\right]\right\}\cos{k_x}+\\
&+2J_2S\cos{k_x}\cos{k_y}\\
\Delta_{\vec{k}2}=&-KS(2S-1)\cos{k_x}\\
\Delta_{\vec{k}m}=&0
\label{eq:delta}
\end{aligned}
\end{equation}





For $m>2$. Finally, the diagonalized Hamiltonian takes the following form.

\begin{equation}
\begin{aligned}
\mathcal{H}_{GSW}=&\sum_{\vec{k}}\left(\epsilon_{\vec{k}1}\beta_{\vec{k}1}^{\dagger}\beta_{\vec{k}1}+\epsilon_{\vec{k}2}\beta_{\vec{k}2}^{\dagger}\beta_{\vec{k}2}\right)+\\
&+\sum_{m=3}^{2S}\epsilon_{\vec{k}m}\beta_{\vec{k}m}^{\dagger}\beta_{\vec{k}m}
\end{aligned}
\end{equation}

with Bogoliubov dispersions:

\begin{equation}
\begin{aligned}
&\epsilon_{\vec{k}1}=2\sqrt{\mu_{\vec{k}1}^2-\Delta_{\vec{k}1}^2}\\
&\epsilon_{\vec{k}2}=2\sqrt{\mu_{\vec{k}2}^2-\Delta_{\vec{k}2}^2}
\end{aligned}
\end{equation}

and flat dispersions: $\epsilon_{\vec{k}m}=2\mu_{\vec{k}m}$ for $m>2$. For convenience, we identify the modes with their value for the z-component of the spin, for the particular case of $S=1$, we will be using. Thus the first mode $m=1$ (with $m_z=0$) becomes $m=0$ and the second mode (with $m_z=-1$) $m=2$ is now represented by $m=\downarrow$.

\section{Dyson--Maleev bosons}
\label{sec:append-DM}

In the methods explained above, the spin operators are expanded around the classical ground state configuration. This approach, however, is not a valid one in the paramagnetic regime, or even in the ordered phase, when the sublattice magnetization becomes too small. To study the temperature evolution of the system, it is thus more appropriate to use an alternate approach, that does not rely on any small parameter. In this representation, the spin operators are expressed in terms of Dyson-Maleev bosons \cite{Dyson1956,Dyson1956a,Maleev1958,Takahashi1989,Takahashi1990}:

\begin{equation}
\begin{aligned}
S^+_{\vec{r}}&=\sqrt{2S}\left(1-\frac{a_{\vec{r}}^{\dagger}a_{\vec{r}}}{2S}\right)a_{\vec{r}} \\
S^-_{\vec{r}}&=\sqrt{2S}a_{\vec{r}}^{\dagger} \\
S^z_{\vec{r}}&=S-a_{\vec{r}}^{\dagger}a_{\vec{r}}
\end{aligned}
\end{equation}

We note that $S_{\vec{r}}^+$ and $S_{\vec{r}}^-$ are no longer complex conjugates. However, since the resulting Hamiltonian is still Hermitian, we are allowed to proceed.

At this point we can treat the biquadratic term using two different approaches. The first one uses the Hubbard-Stratonovich procedure outlined above, where the higher order term is substituted by the following decoupling: $\left(\vec{S}_{\vec{r}}\cdot\vec{S}_{\vec{r'}}\right)^2\simeq 2\left<\vec{S}_{\vec{r}}\cdot\vec{S}_{\vec{r'}}\right>\vec{S}_{\vec{r}}\cdot\vec{S}_{\vec{r'}}-\left<\vec{S}_{\vec{r}}\cdot\vec{S}_{\vec{r'}}\right>^2$. The results obtained by this method have already been studied for the finite temperature case in Ref.~\onlinecite{Yu2012}. An alternate approach is to decouple the entire biquadratic term using all possible decouplings via Wick's theorem and the averages specified before. This method was used by Ref.~\onlinecite{Holt2011} in the $T=0$ case and we expand it to include the finite temperature regime. Once again, we obtain dispersions of the form: $\omega_{\vec{k}}=2\sqrt{A_{\vec{k}}^2-B_{\vec{k}}^2}$ for the two- and three-dimensional cases, respectively. Following the notation from Ref.~\onlinecite{Holt2011},

\begin{equation}
\begin{aligned}
A_{\vec{k}(2D)}=&\lambda+2J_2(S-\alpha_{xy})+J_1\left[r_x(S-\alpha_x)+\right.\\
&\left.+r_y(S-\beta_y)(\cos{k_y}-1)\right]\\
B_{\vec{k}(2D)}=&J_1r_x(S-\alpha_x)\cos{k_x}+\\
&+2J_2(S-\alpha_{xy})\cos{k_x}\cos{k_y}
\end{aligned}
\end{equation}

\begin{equation}
\begin{aligned}
A_{\vec{k}(3D)}=&A_{\vec{k}(2D)}+J_c(S-\alpha_z)\\
B_{\vec{k}(3D)}=&B_{\vec{k}(2D)}+J_c(S-\alpha_z)\cos{k_z}
\end{aligned}
\end{equation}

Where $r_x$, $r_y$ stand for the following expressions:

\begin{equation}
\begin{aligned}
r_x=&1+\frac{K}{S-\alpha_x}\left[2S^3-2S^2(1+5\alpha_x)+\right.\\
&\left.+S(18\alpha_x^2+8\alpha_x+1)-12\alpha_x^3-9\alpha_x^2-2\alpha_x\right]\\
r_y=&1-\frac{K}{S-\beta_y}\left[2S^3-2S^2(1+5\beta_y)+\right.\\
&\left.+S(18\beta_y^2+8\beta_y)-12\beta_y^3-9\beta_y^2-\beta_y\right]
\end{aligned}
\end{equation}

Finally, we introduced the chemical potential $\lambda$ to enforce the constraint $\left<S_z\right>=0$ in the paramagnetic regime, which results in the following equation:

\begin{equation}
\begin{aligned}
S=&\frac{1}{N_s}\sum_{\vec{k}}\left<a_{\vec{k}}^{\dagger}a_{\vec{k}}\right>\rightarrow S+\frac{1}{2}=\frac{1}{N_s}\sum_{\vec{k}}\left(1+2n_{\vec{k}}\right)\frac{A_{\vec{k}}}{\omega_{\vec{k}}}
\end{aligned}
\end{equation}

Thus, $\lambda=0$ in the magnetically ordered phase. The resulting set of self-consistent equations has the same form of (\ref{eq:system}), with the addition of the following equation in the three-dimensional case:

\begin{equation}
\begin{aligned}
\alpha_z=&\frac{1}{N_s}\sum_{\vec{k}}\left(\left<a_{\vec{k}}^{\dagger}a_{\vec{k}}\right>+\left<a_{\vec{k}}a_{\vec{-k}}\right>\cos{k_z}\right)=\\
&=-\frac{1}{2}+\frac{1}{N_s}\sum_{\vec{k}}\left(1+2n_{\vec{k}}\right)\frac{A_{\vec{k}}-B_{\vec{k}}\cos{k_z}}{\omega_{\vec{k}}}
\end{aligned}
\end{equation}

We differentiate the two domains, $T<T_N$ (where $n<S$ and $\lambda=0$) and $T>T_N$ (where $n=S$ and $\lambda\neq0$).

\end{document}